\documentclass[reprint,prb,a4paper,amsfonts,amssymb,floats,titlepage,fleqn,lengthcheck,twocolumn,
tightenlines,10pt,showpacs,floatfix,superscriptaddress,footinbib,balancelastpage, hidelinks,longbibliography]{revtex4-2}%citeautoscript 
\usepackage{framed}
\usepackage[utf8]{inputenc}
\usepackage[T1]{fontenc}
\usepackage[version=4]{mhchem}
\usepackage{comment}
\newcommand{\Onlinecite}[1]{\hspace{-1 ex} \nocite{#1}\citenum{#1}} 
\usepackage{subfiles}
\usepackage{mathastext}
\usepackage{amsmath}
\usepackage{amssymb}
\usepackage[nice]{nicefrac}
\usepackage{amsthm}
\usepackage{mathtools}
\setcitestyle{super}
\usepackage{xcolor}
\usepackage[normalem]{ulem}
\usepackage{nicematrix}
\usepackage{enumitem}
\usepackage{titlesec}
\usepackage{hyperref}
\usepackage{xr}
\usepackage{natbib}
\begin{document}
%\title{Quantum-Critical Residual Resistivity and Emergent Universal Correlations in the Fluctuation-Driven Fermi-Liquid Regime of Heavy-Fermion Superconductors}
\title{Quantum-Critical, Spin-Fluctuation-driven Residual Resistivity and Emergent Universal Correlations in the Fermi-Liquid Regime of Heavy-Fermion Superconductors}
\author{M. ElMassalami}
\email  {massalam@if.ufrj.br}
\affiliation{Instituto de F\'{\i}sica, Universidade Federal do Rio de Janeiro, Caixa Postal 68528, 21941-972 Rio de Janeiro RJ, Brazil}
\author{P. B. Castro}
%\email  {CASTRO.Pedro@nims.go.jp}
\affiliation{Instituto de F\'{\i}sica, Universidade Federal do Rio de Janeiro, Caixa Postal 68528, 21941-972 Rio de Janeiro RJ, Brazil}
\author{M. B. Silva Neto}
\email {mbsn@if.ufrj.br}
\affiliation{Instituto de F\'{\i}sica, Universidade Federal do Rio de Janeiro, Caixa Postal 68528, 21941-972 Rio de Janeiro RJ, Brazil}
%
%\keywords{Keyword1, Keyword2, Keyword3} 
%
\begin{abstract}
We investigate correlations within the unconventional Fermi-liquid (FL) regime of quantum-critical (QC) heavy-fermion superconductors by tracking the pressure dependence of three quantities: the temperature-independent, spin-fluctuation-driven residual resistivity, $\rho^{\mbox{\tiny sf}}_{\mbox{\tiny 0}}(P)$; the FL scattering coefficient, $A(P)$; and the superconducting transition temperature, $T_c(P)$. The first two define the spin-fluctuation contribution to the resistivity, $\rho(T)=\rho^{\mbox{\tiny sf}}_{\mbox{\tiny 0}}+AT^2$.
Using experimental data from archetypal heavy-fermion systems, we identify three robust empirical correlations: $\ln(\nicefrac{T_c}{\theta}) \propto A^{-1/2}$, $A \propto (\rho^{\mbox{\tiny sf}}_{\mbox{\tiny 0}})^2$, and $\ln(\nicefrac{T_c}{\theta}) \propto \big(\rho^{\mbox{\tiny sf}}_{\mbox{\tiny 0}}\big)^{-1}$ ($\theta$ is a characteristic temperature scale). Absent in conventional FL superconductors, these relationships indicate that QC fluctuations not only mediate inelastic scattering and Cooper pairing, but also generate an effective elastic channel responsible for $\rho^{\mbox{\tiny sf}}_{\mbox{\tiny 0}}$. We explicitly calculate $\rho^{\mbox{\tiny sf}}_{\mbox{\tiny 0}}$ on the high-pressure side of the quantum critical point (QCP) and introduce a characteristic length scale, $\ell \sim \big(\rho^{\mbox{\tiny sf}}_{\mbox{\tiny 0}}\big)^{-1}$, that captures the spatial extent of fluctuation-induced scattering. Within this regime, and within the Migdal--Eliashberg framework combined with Boltzmann transport theory, we derive analytic expressions for $T_c(\ell)$ and $A(\ell)$, together with their interrelations, which are consistent with the observed empirical trends.
These findings highlight the quantum-critical FL regime in heavy-fermion superconductors as an intrinsically correlated phase, governed by fluctuations and marked by unconventional transport and pairing mechanisms.
\end{abstract}
\maketitle
%\linenumbers
\section{Introduction \label{Sec.Introduction}}
In the study of unconventional superconductors—including heavy-fermion, cuprate, and iron-based systems—a central question concerns the microscopic mechanism that binds electrons into Cooper pairs. In pursuit of answers, extensive efforts have focused on identifying universal correlations that link the onset or characteristics of superconductivity to those of the complex, often competing, cascade of adjacent or intertwined normal-state phases.
%—especially within regions of the phase diagram where these phases are proximate or intertwined. 
This search for correlations is usually conducted across \textit{T–X} phase diagrams that span a broad range of tuning parameters \textit{X}, such as pressure (\textit{P}), chemical substitution (\textit{x}), or magnetic field (\textit{H}). Two of the most extensively studied \textit{T–X} phase diagrams are those of heavy-fermion (HF) systems\cite{Gegenwart08-QC-HF-metals,Coleman07-HF-Review,Stewart84-HeavyFermion-Review,Jaccard99-HF-SC-Mag-HP,Yuan03-CeCu2Si2-SC-Phases-HF,Yang14-SC-Heavy-Electrons} and 
high-\textit{T}$_c$ cuprates.\cite{Norman11-Unconventional-SUC-Challenge,Scalapino12-UnconSCs-CommonThread,Pines13-PhaseDiagramCuprates,Keimer15-Cuprate-Quantum-Matter-HTC-SUC,Stewart17-Unconventional-SUC-Review}
Within the low-temperature \textit{P–T} phase diagram of quantum-critical heavy-fermion (QCHF) superconductors—the sole focus of this work—the tuning parameter $P$ stabilizes a sequence of phases:\cite{Gegenwart08-QC-HF-metals,Coleman07-HF-Review,Stewart84-HeavyFermion-Review,Jaccard99-HF-SC-Mag-HP,Yuan03-CeCu2Si2-SC-Phases-HF,Yang14-SC-Heavy-Electrons}
starting from a metallic antiferromagnet (AFM), passing through a non-Fermi-liquid (NFL) regime with anomalous transport (e.g., linear-in-temperature resistivity) near a quantum critical point (QCP) at $P_c$, followed by the emergence of superconductivity, and eventually reaching a Fermi-liquid (FL) phase at pressures beyond $P_c$.

The analytical search for informative correlations in QCHF systems is most naturally carried out within the Kondo-lattice framework, which models the interaction between localized magnetic moments and conduction electrons as:\cite{Yang17-QC-Scaling-FLuctuation-Kondo-Lattice}
\begin{widetext}
\begin{equation}
\begin{aligned}
H_{\mathrm{int}}
&= J \sum_i \mathbf{S}_i \cdot \mathbf{s}_c(\mathbf{r}_i)
= \frac{J}{2}\sum_{i,\alpha\beta} c_{i\alpha}^{\dagger}\,\boldsymbol{\sigma}_{\alpha\beta}\,c_{i\beta}\cdot \mathbf{S}_i 
&= \frac{J}{2N}\sum_{\mathbf{k},\mathbf{q}} \sum_{\alpha\beta}
c_{\mathbf{k}+\mathbf{q},\alpha}^{\dagger}\,\boldsymbol{\sigma}_{\alpha\beta}\,c_{\mathbf{k}\beta}\cdot \mathbf{S}_{-\mathbf{q}} ,
\end{aligned}
\label{Eq-Kondo-Lattice-Hamiltonian}
\end{equation}
\end{widetext}
where $J$ is the Kondo coupling, $\mathbf{S}_i$ is the spin of the localized $f$ electron at site $i$, and $\mathbf{s}_c(\mathbf{r}_i) = \tfrac{1}{2}\, c^\dagger_{\alpha}(\mathbf{r}_i)\,\boldsymbol{\sigma}_{\alpha\beta}\,c_{\beta}(\mathbf{r}_i)$ is the spin density of conduction electrons, with Pauli matrices $\boldsymbol{\sigma}$.

These investigations are most naturally carried out on the high-pressure side of quantum criticality, where superconductivity lies adjacent to a correlated Fermi-liquid (FL) normal state. In this regime, control parameters modulate the interaction between quasiparticles and their medium, enabling a direct assessment of how fluctuations govern pairing and transport: e.g., (i) renormalize quasiparticle properties—affecting motion, energy, and lifetime—and (ii) reshape the host matrix—governing the emergence and stability of normal and superconducting phases. Together, they define the structure and boundaries of this selected portion of the phase diagram. 

An effective molecular-field–type interaction can be introduced, characterized by a constant $g_m$, through which quasiparticle spins $\mathbf{s}_1$ and $\mathbf{s}_2$ couple to magnetic fluctuations of the matrix described by a dynamical susceptibility $\chi_m(\mathbf{q},\omega)$:
\cite{Mathur98-HF-SC-Mag-Mediation,Monthoux07-SC-Without-Phonon}
%\begin{widetext}
\begin{eqnarray}
V^{\mathrm{sf}}_{ee}(\mathbf{q},\omega)
= -\frac{J^2}{4}\,\chi_f(\mathbf{q},\omega)\,\boldsymbol{\sigma}_1\!\cdot\!\boldsymbol{\sigma}_2 &\nonumber\\
= g_m^2\,\chi_m(\mathbf{q},\omega)\,(\mathbf{s}_1\!\cdot\!\mathbf{s}_2)&
=\mathbf{I}^2\,\chi(\mathbf{q},\omega),
\label{Eq.Effective-Interaction-Potential}
\end{eqnarray}
%\end{widetext}
where $\chi_f$ is the spin susceptibility and $\mathbf{I}$ denotes an effective exchange coupling.

Within this framework, two key quantities, the scattering rate, $\Gamma$, and the coupling constant, $\lambda$, are central for identifying correlations and delimiting regions of the phase diagram, including the quantum-critical region with its unconventional, strongly correlated states. 

The scattering rate $\Gamma(T,\omega)$ quantifies quasiparticle longevity and momentum/energy relaxation. Its evolution mirrors the \textit{P–T} phase diagram of QCHF superconductors: in the low-pressure antiferromagnetic (AFM) phase, scattering is relatively weak and dominated by low-energy inelastic processes; on approaching the QCP, enhanced Quantum-critical fluctuations (QCFs) drive a non-Fermi-liquid (NFL) regime with $\Gamma(T,\omega)\propto \max(\omega,T)$, producing linear-in-$T$ and/or linear-in-$\omega$ behavior; beyond the critical pressure ($P>P_c$), well-defined quasiparticles are restored and $\Gamma(T,\omega)\propto \omega^2+(\pi k_B T)^2$, as expected for a correlated FL phase. 

The coupling constant $\lambda(T,\omega)$, on the other hand, measures the strength of electronic coupling to bosonic excitations and controls the effective mass, band-structure renormalization, and the superconducting scale $T_c$. Across the high-P side of the \textit{P–T} phase diagram, $\lambda$ is strongly enhanced near the QCP and decreases away from it; correspondingly, parameters such as effective mass $m^{\star}$, specific-heat coefficient $\gamma$, and—absent strong pair breaking—$T_c$ exhibit the same qualitative trend.

More generally, the correlated FL regime emphasized in this work is not a simple conventional background state, but rather a fluctuation-shaped heavy-Fermi-liquid regime. In it, the same external tuning parameter that controls the distance to a fluctuation-rich region also governs the renormalization quantities $\lambda$, $\gamma$, $m^{\star}$, and $A$. Approaching a QCP, phase boundary, or crossover regime from the FL side enhances the relevant quantum-critical fluctuations and, with them, the quasiparticle renormalization; moving away weakens all these quantities together. Within this framework, $m^{\star}$ is indeed one of the clearest indicators of QCFs, but not an isolated one: it is part of a broader and internally consistent set of heavy-fermion renormalization parameters that track the same fluctuation physics.

For the empirical mapping of these parameters and \textit{P–T} phase diagrams, multiple probes are available. Among them, conventional \textit{P}- and \textit{T}-dependent resistivity is particularly convenient for the present study because it simultaneously enables (i) the search for QC-fluctuation–driven correlations and (ii) the identification and analysis of phase cascades and their boundaries within the corresponding \textit{P–T} diagram.
Here, our search for correlations is best served by investigating how QCFs shape the baric and thermal evolution of the resistivity in QCHF superconductors. This influence is captured, as shown below, by three low-energy parameters readily extracted from the low-$T$ resistivity curves: (i) the superconducting critical temperature $T_c$ (a fingerprint of QC-fluctuation–mediated pairing); (ii) the coefficient of the {\it square-in-$T$} term, $A^{\mbox{\tiny sf}}$ (a fingerprint of QC-fluctuation–driven inelastic scattering); and (iii) an excess residual resistivity $\rho^{\mbox{\tiny sf}}_{\mbox{\tiny 0}}$ (a fingerprint of an effective elastic channel arising from slow, long-wavelength critical modes).
Particularly informative is the baric evolution of the resistivity in the low-$T$ high-pressure side of the QCP, where QCFs can persist as $T\!\to\!0$, consistent with the uncertainty principle. Because the effective density of fluctuation modes varies with pressure, $\rho^{\mbox{\tiny sf}}_{\mbox{\tiny 0}}(P)$ is expected to peak in the same pressure window where $T_c(P)$ is maximal. This expectation motivated our search for correlations among $\rho^{\mbox{\tiny sf}}_{\mbox{\tiny 0}}(P)$, $A(P)$, and $T_c(P)$.
%
%%%%%%%%%%%%%%%%%% Begin Fig  1  %%%%%%%%%%
\begin{figure}
\includegraphics[scale=0.25]{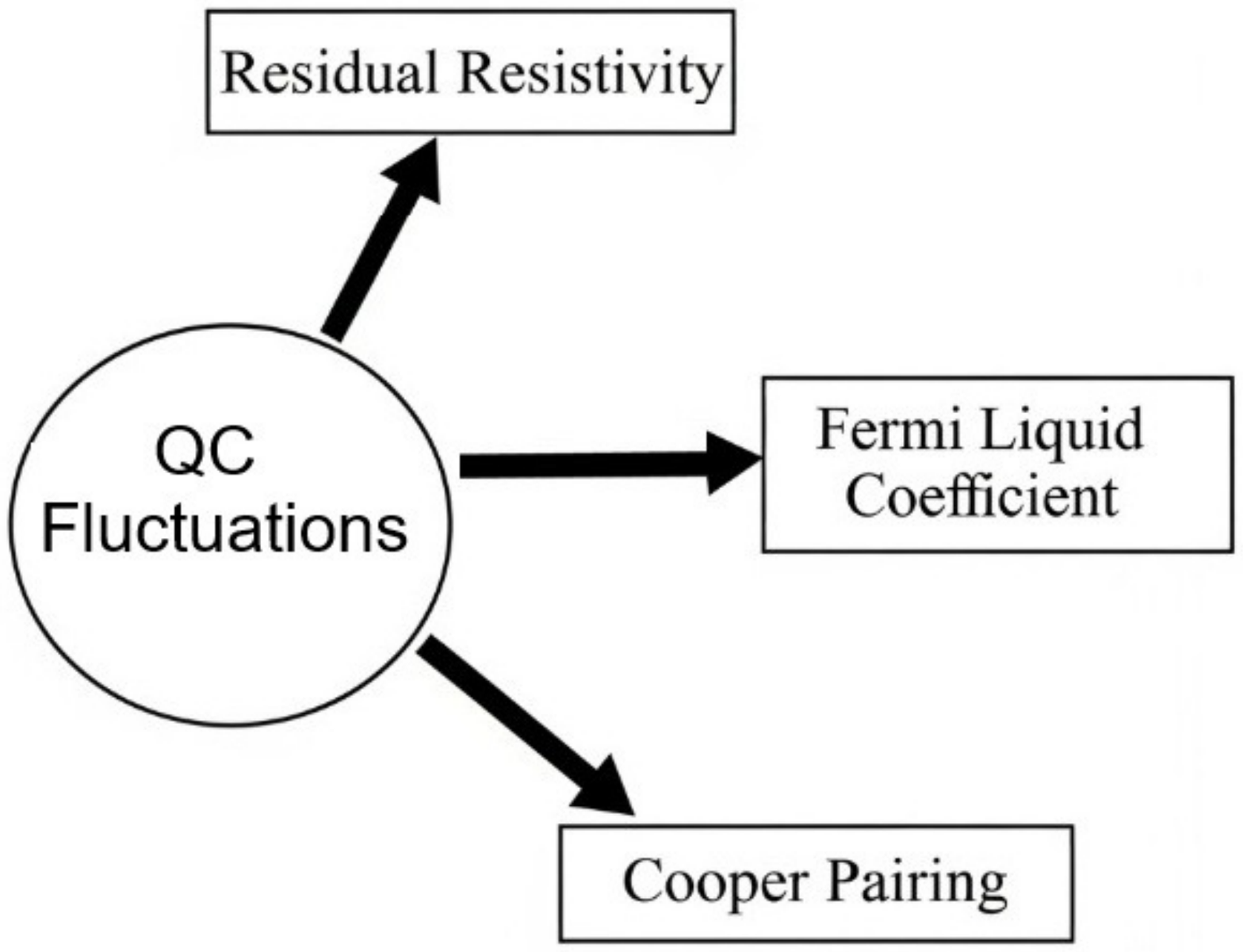}%
\caption{
The threefold role for QCFs within the fluctuation-driven FL regime of QCHF superconductors: (1) they mediate the effective pairing interaction that sets $T_c$; (2) they generate inelastic quasiparticle scattering that controls $A^{sf}$; and (3) they contribute an \emph{effective elastic} channel that yields a finite residual resistivity, $\rho^{\mbox{\tiny sf}}_{\mbox{\tiny 0}}$, analogous in its transport signature to static disorder yet rooted in critical modes rather than quenched impurities. The emergence of these three effects, together with the empirical inter-relations, points to a unified fluctuation mechanism governing pairing, transport, and residual scattering, all studied within this particular side of the QCP.
}
\label{Fig1-Spin_Fluctuations_Rho0_Tc_A}
\end{figure}
%%%%%%%%%%%%%%%%%%%   End Fig  1  %%%%%%%%%%%%%%%%%%%%%%%%%%%

Theoretically, we show that these QC-fluctuation–related quantities $\rho^{\mbox{\tiny sf}}_{\mbox{\tiny{0}}}$, $A^{sf}$, and $T_c$ are governed by the same $\Gamma(T,\omega)$ and $\lambda(T,\omega)$, which in turn follow from the Kondo-lattice effective interaction $V^{\mathrm{sf}}_{ee}(\mathbf q,\omega)$ and the magnetic susceptibility $\chi_m(\mathbf q,\omega)$. Demonstrating combined empirical and theoretical correlations among $\rho^{\mbox{\tiny sf}}_{\mbox{\tiny{0}}}$, $A^{sf}$, and $T_c$ thereby establishes that quantum-critical fluctuations simultaneously (i) mediate Cooper pairing (encoded in $T_c$), (ii) drive inelastic transport (encoded in $A^{sf}$), and (iii) produce an effective elastic channel (encoded in $\rho^{\mbox{\tiny sf}}_{\mbox{\tiny{0}}}$) (see Fig.\ref{Fig1-Spin_Fluctuations_Rho0_Tc_A}). Related studies have already explored the $T_c$–$A$ correlation in optimally doped and overdoped high-$T_c$ superconductors,\cite{Wakimoto04-HTC-Overdoped-Low-Excitation-SUC-Correlations,Taillefer10-Scattering-pairing-HTc-Cuprate,Greene20-Strange-Metal-Electron-doped,Maier20-Overdoped-End-Cuprates-PhaseDiagram,Yuan22-Scaling-Strange-Metal-Cuprates} including an empirical scaling between $T_c$ and the coefficient of a {\it linear-in-$T$} term, $A_{\mathrm{NFL}}$, within the NFL regime of cuprates.\cite{Yuan22-Scaling-Strange-Metal-Cuprates}

The following text is organized as follows. In \S\ref{Sec.Emprical-Derivation}, we analyze the resistivity of three archetypal QCHF superconductors to extract and generalize three empirical correlations among $\rho^{\mbox{\tiny sf}}_{\mbox{\tiny{0}}}(P)$, $T_c(P)$, and $A(P)$. In \S\ref{Sec.Theoretical-Derivation}, starting from the Kondo-lattice model and applying Migdal–Eliashberg theory together with Boltzmann transport, we derive analytic expressions for $\rho^{\mbox{\tiny sf}}_{\mbox{\tiny{0}}}$, $T_c$, and $A$, as well as their mutual correlations, and show agreement with experimental trends. The Appendices contrast the  FL state of a QCHF superconductor with that of a conventional FL, and illustrates the broader reach of our approach by deriving the Kadowaki–Woods relation and the gap-to-$T_c$ ratio for QCHF superconductors.
%
%%%%%%%%%%%%%%%%%% Begin Fig  2  %%%%%%%%%%[trim={<left> <lower> <right> <upper>}
\begin{figure*}[hbtp]
\centering
\includegraphics[scale=0.5,trim={0cm 0cm 0cm 0cm}, clip]{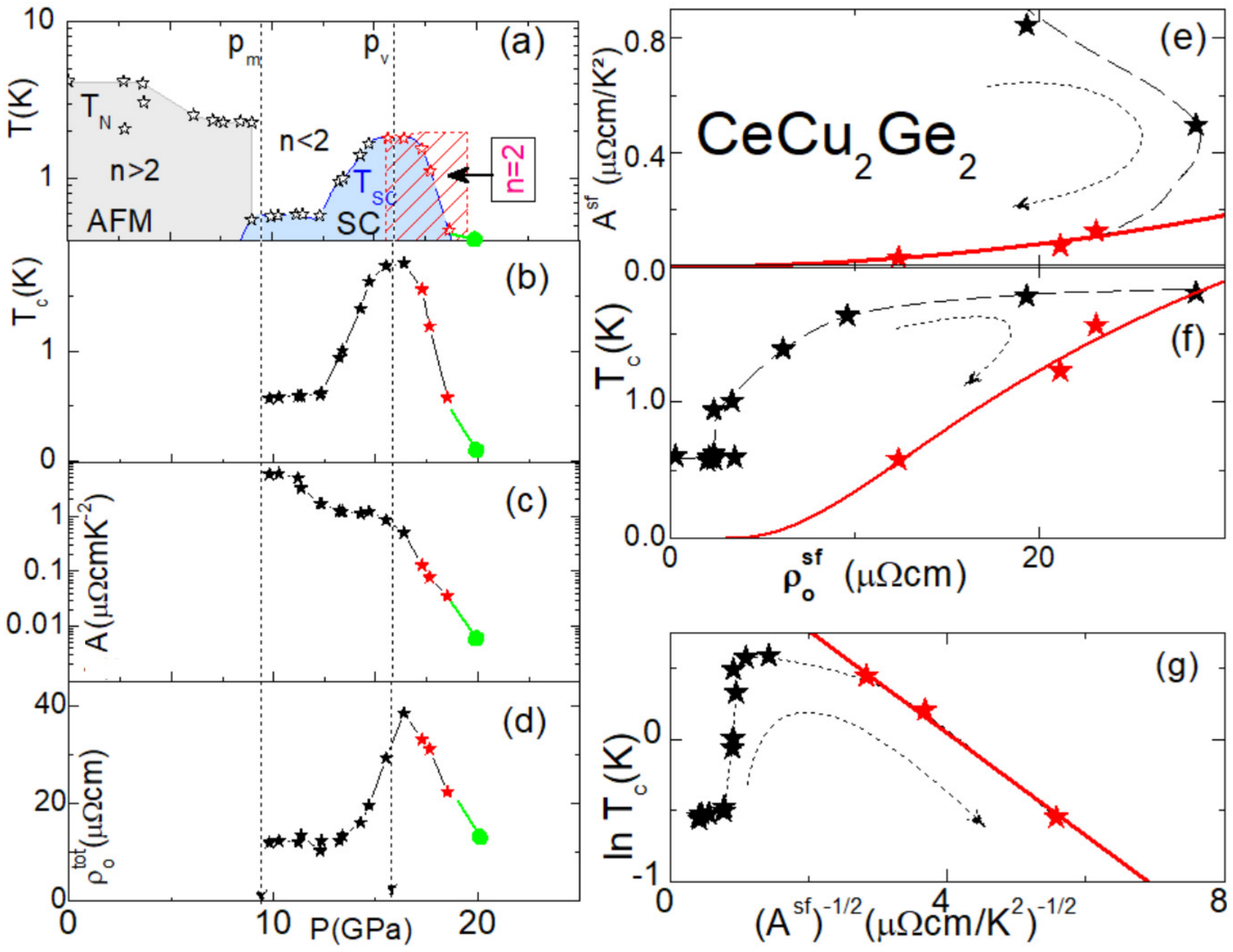}%
\caption{ \textit{T-P} phase diagram and baric evolution of the parameters of \ce{CeCu2Ge2} (datasets taken from Ref.\,\protect\Onlinecite{Jaccard99-HF-SC-Mag-HP}).  \textbf{(a)}  A semi-log \textit{T-P} phase diagram showing the evolution of $T_{N}(P)$, $T_{c}(P)$, and the exponent \textit{n}. Vertical dashed arrows represent $P_{m} \approx 9.4\,$ GPa and $P_{v} \approx 16\,$ GPa.\cite{Jaccard99-HF-SC-Mag-HP} The red symbols correspond to the QCF-driven FL regime, shown by the hatched area and terminating at the solid green circle. The dashed curves are visual guides. Baric evolution of \textbf{(b)}  $T_{c}(P)$, \textbf{(c)}  $A(P)$ in a semi-log plot, and 
\textbf{(d)}  $\rho^{\mbox{\tiny{tot}}}_{\mbox{\tiny{0}}}(P)$.  
\textbf{(e)}  Correlation between $A^{\mbox{\tiny sf}}$ and $\rho^{\mbox{\tiny sf}}_{\mbox{\tiny 0}}(P)$. The red curve shows the fit to Eq.\ref{Eq.A-vs-rho} with $A_1\!\approx\!0$; the resulting parameters $A_0$ and $A_2$ are listed in Table\,\ref{Tab.Fit-Values-A-Ro-Theta-F}.
\textbf{(f)}  $T_c$ \textit{versus} $\rho^{\mbox{\tiny sf}}_{\mbox{\tiny 0}}$. The red curve is a fit to Eq.\ref{Eq.lnTc-vs-ro-one}a. 
\textbf{(g)}  $\ln T_c$ \textit{versus} $A^{-1/2}$. The red line is a fit to the linearized Eq.\ref{Eq.Tc-of-A}b; the fit parameters $\theta$ and $\mathcal{F}$ are listed in Table\,\ref{Tab.Fit-Values-A-Ro-Theta-F}. 
}\label{Fig2-CeCu2Ge2}
\end{figure*}
%%%%%%%%%%%%%%%%%%%   End Fig  2  %%%%%%%%%%%%%%%%%%%%%%%%%%%
\section{Empirical correlations among \texorpdfstring{$\rho^{\mbox{\tiny{sf}}}_{\mbox{\tiny{0}}}$, $A$, and \textit{T$_{c}$}}{} \label{Sec.Emprical-Derivation} }
%\subsection{Empirical correlations among \texorpdfstring{$\rho^{\mbox{\tiny{sf}}}_{\mbox{\tiny{0}}}$, $A$, and \textit{T$_{c}$}}{} \label{Sec.Emprical-Derivation}}
%
Below, we analyze the reported baric and thermal evolution of $\rho^{\mbox{\tiny{tot}}}(T,P)$ for three representative QCHF superconductors. These compounds were selected because the literature provides comprehensive \textit{P--T} phase diagrams and high-quality resistivity datasets, enabling reliable extraction of $\rho^{\mbox{\tiny sf}}_{\mbox{\tiny 0}}(P)$, $A(P)$, and $T_c(P)$, as well as of the correlations among these quantities.

\subsection{\texorpdfstring{\textrm{\ce{CeCu2X2} ($X$=Si, Ge)}}{} \label{SubSec.CeCu2X2}}
\begin{figure*}[hbtp]
\centering%
\includegraphics[scale=0.5,trim={0cm 0cm 0cm 0cm}, clip]{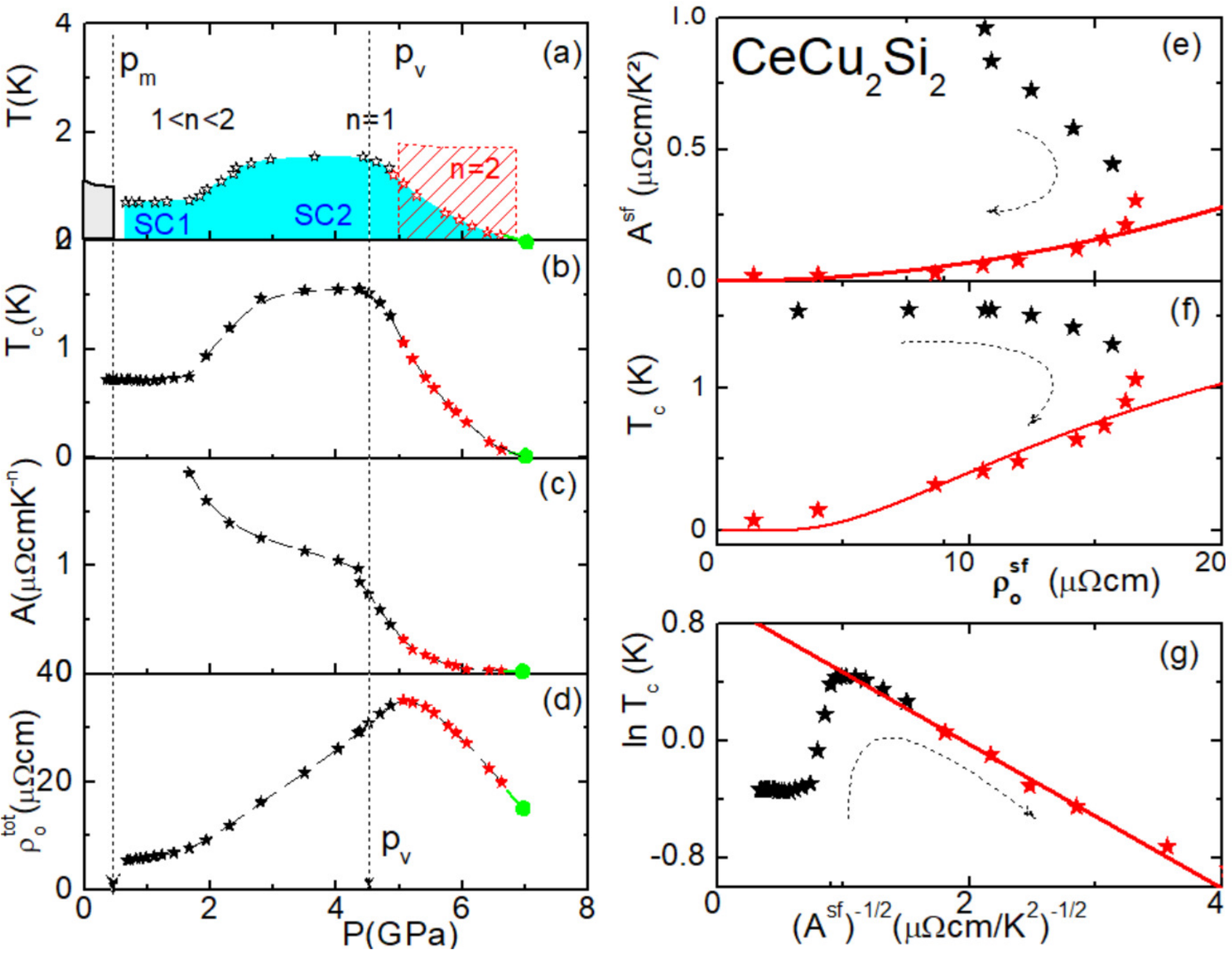}%
\caption{ \textbf{(a)}  The  \textit{T-P} phase diagram and baric evolution of  \textbf{(b)}  $T_{c}(P)$,  \textbf{(c)}  $A(P)$, and \textbf{(d)}  $\rho^{\mbox{\tiny{tot}}}_{\mbox{\tiny{0}}}(P)$ of \ce{CeCu2Si2}. All datasets  were taken from the continuous line (used there as a visual guide) of Ref.\,\protect\Onlinecite{Holmes04-CeCu2Si2-ValenceFluctuation}. The red symbols belong to the QCF-driven FL regime  while the dashed lines are visual guides.  It is noted that, due to non-stoichiometry, defects, impurities or disorder, the reported parameters of \ce{CeCu2Si2} vary considerably from sample to sample; nonetheless, the overall evolution as well as the derived correlations are consistently similar.\cite{Rosch99-disorder-SpinFluctuation-QCP,Kambe97-CeCu2Si2-Purity,Sheikin00-CeCu2Si2-HF-R-P,Gegenwart98-CeCu2Si2-Brakup-HF} %
\textbf{(e)}  Correlation between $A^{\mbox{\tiny sf}}$ and $\rho^{\mbox{\tiny sf}}_{\mbox{\tiny 0}}(P)$. The red curve shows the fit to Eq.\ref{Eq.A-vs-rho} with $A_1\!\approx\!0$; the resulting parameters $A_0$ and $A_2$ are listed in Table\,\ref{Tab.Fit-Values-A-Ro-Theta-F}.
\textbf{(f)}  $T_c$ \textit{versus} $\rho^{\mbox{\tiny sf}}_{\mbox{\tiny 0}}$. The red curve is a fit to Eq.\ref{Eq.lnTc-vs-ro-one}a. 
\textbf{(g)}  $\ln T_c$ \textit{versus} $A^{-1/2}$. The solid red line is a fit to the linearized Eq.\ref{Eq.Tc-of-A}b; the fit parameters $\theta$ and $\mathcal{F}$ are listed in Table\,\ref{Tab.Fit-Values-A-Ro-Theta-F}. 
}%
\label{Fig3-CeCu2Si2}
\end{figure*}

The \textit{T–P} phase diagrams of \ce{CeCu2Ge2} and \ce{CeCu2Si2} are shown in Fig.\ref{Fig2-CeCu2Ge2}(a) [Refs.\,\Onlinecite{Jaccard99-HF-SC-Mag-HP,Honda13-CeCu2Ge2-Mag-SC}] 
and Fig.\ref{Fig3-CeCu2Si2}(a) [Refs.\,\Onlinecite{Yuan03-CeCu2Si2-SC-Phases-HF,Gegenwart98-CeCu2Si2-Brakup-HF,Bellarbi84-CeCu2Si2-Valence-Instability,Rueff11-CeCu2Si2-VQCP,Holmes04-CeCu2Si2-ValenceFluctuation}], respectively. 
The pressure dependence of the extracted $T_c$, $A$, and $\rho^{\mbox{\tiny tot}}_{\mbox{\tiny{0}}}$ is shown in Figs.\ref{Fig2-CeCu2Ge2}(b–d) and \ref{Fig3-CeCu2Si2}(b–d), displaying pronounced variations near the critical pressure. These features agree with the argument in the Introduction (see also \S\ref{SubSec.QC-Fluctuation-Role}) that QCFs are enhanced on approaching the QCP and diminish progressively away from it.
Remarkably, the coefficient $A$ in the FL regime of compounds such as \ce{CeCu2X2} exceeds, by more than five orders of magnitude, the value expected for conventional FL metals ($A_{ee}^{h} \simeq 10^{-7} \mu\Omega\text{cm}/\text{K}^2$).\cite{Lawrence73-ee-Scattering-in-Res, Lawrence76-ee-Scattering-NobleMetals, MacDonald80-e-F-enhanced-e-e-Interaction, MacDonald81-Umkalpp-resistivity, Patton75-FermiLiquid-Superfluid, Pethick86-FL-Theory-UPt3}
A similarly anomalous enhancement is observed in the residual resistivity $\rho^{\mbox{\tiny{sf}}}_{\mbox{\tiny{0}}}$ when compared with typical FL metals. These two features—together with the emergence of superconductivity adjacent to the FL regime—underscore the distinct and nontrivial character of the QCF-driven FL state in these QCHF superconductors (see also Appendix \ref{Sec.A-Comparison-FLs}).
Furthermore, a strong correlation is evident in the pressure evolution of $T_c$, $A$, and $\rho^{\mbox{\tiny sf}}_{\mbox{\tiny 0}}$ within this QCF-driven FL regime, highlighted by the red symbols inside the red-hatched regions in Figs.\ref{Fig2-CeCu2Ge2}(b–d) and \ref{Fig3-CeCu2Si2}(b–d).
In all cases considered here, this regime begins at $P_L$, where the resistivity exponent reaches $n=2$, and ends at $P_H$ (marked by the solid green circle), where extrapolations give $T_c \to 0$, $A^{\mbox{\tiny tot}} \to A_{\mbox{\tiny 0}}$, and $\rho^{\mbox{\tiny tot}}_{\mbox{\tiny 0}} \to \rho^{\mbox{\tiny 0}}_{\mbox{\tiny 0}}$, with $A_{\mbox{\tiny 0}}$ and $\rho^{\mbox{\tiny 0}}_{\mbox{\tiny 0}}$ denoting the strongly weakened QCF contribution together with possible non-QCF-related background terms.

To empirically probe correlations among the key quantities, we examine the following parametric plots:
(i) $A^{\mbox{\tiny sf}}(P)$ vs.\ $\rho^{\mbox{\tiny sf}}_{\mbox{\tiny 0}}(P)$ [Figs.~\ref{Fig2-CeCu2Ge2}(e), \ref{Fig3-CeCu2Si2}(e)], where $A^{\mbox{\tiny sf}}\equiv A^{\mbox{\tiny tot}}-A_{\mbox{\tiny 0}}$ and $\rho^{\mbox{\tiny sf}}_{\mbox{\tiny 0}}(P)\equiv \rho^{\mbox{\tiny tot}}_{\mbox{\tiny 0}}(P)-\rho^{\mbox{\tiny 0}}_{\mbox{\tiny 0}}$;
(ii) $T_c(P)$ vs.\ $\rho^{\mbox{\tiny sf}}_{\mbox{\tiny 0}}(P)$ [Figs.~\ref{Fig2-CeCu2Ge2}(f), \ref{Fig3-CeCu2Si2}(f)];
(iii) $\ln T_c(P)$ vs.\ $\big(A^{\mbox{\tiny sf}}(P)\big)^{-1/2}$ [Figs.~\ref{Fig2-CeCu2Ge2}(g), \ref{Fig3-CeCu2Si2}(g)].
From these plots—within the QCF-driven FL regime on the FL side of $P_c$—three noteworthy correlations emerge:
\begin{itemize}[label=$\star$]%[label=$\diamond$]
\item $A^{\mbox{\tiny sf}}(P)$ vs.\ $\rho^{\mbox{\tiny sf}}_{\mbox{\tiny 0}}(P)$:
the variation of $A^{\mbox{\tiny sf}}$ with $\rho^{\mbox{\tiny sf}}_{\mbox{\tiny 0}}$ follows a quadratic dependence,\cite{Gegenwart08-QC-HF-metals,Coleman07-HF-Review,Miranda05-disorder-driven-NFL-Review,Sheikin00-CeCu2Si2-HF-R-P,Gegenwart98-CeCu2Si2-Brakup-HF}
\begin{equation}
A^{\mbox{\tiny sf}} \equiv A^{\mbox{\tiny tot}}-A_{\mbox{\tiny 0}}
\;\approx\;
A_2\,\big(\rho^{\mbox{\tiny sf}}_{\mbox{\tiny 0}}\big)^2.
\label{Eq.dominant-A-defect}
\end{equation}
The absence of a linear-in-$\rho^{\mbox{\tiny sf}}_{\mbox{\tiny 0}}$ term rules out both an NFL contribution and a Koshino–Taylor mechanism.
\item $T_c(P)$ vs.\ $\rho^{\mbox{\tiny sf}}_{\mbox{\tiny 0}}(P)$:
$T_c$ depends \textit{exponentially} on $1/\rho^{\mbox{\tiny sf}}_{\mbox{\tiny 0}}$.
\item $T_c(P)$ vs.\ $A^{\mbox{\tiny sf}}(P)$:
$\ln T_c$ is \textit{linear} in $\big(A^{\mbox{\tiny sf}}\big)^{-1/2}$.
\end{itemize}
%%%%%%%%%%%%%%%%%%% Begin  Fig3  %%%%%%%%%%%%%%%%%%%%%%%%%%%
%%%%%%%%%%%%%%%%%%% End  Fig  3  %%%%%%%%%%%%%%%%%%%%%%%%%%%
%%%%%%%%%%%%%%%%%% Begin Fig  4  %%%%%%%%%%%%%%%%%%%%%%%%%%%
\begin{figure*}[hbtp]
\centering%
\includegraphics[scale=0.5,trim={0cm 0cm 0cm 0cm}, clip]{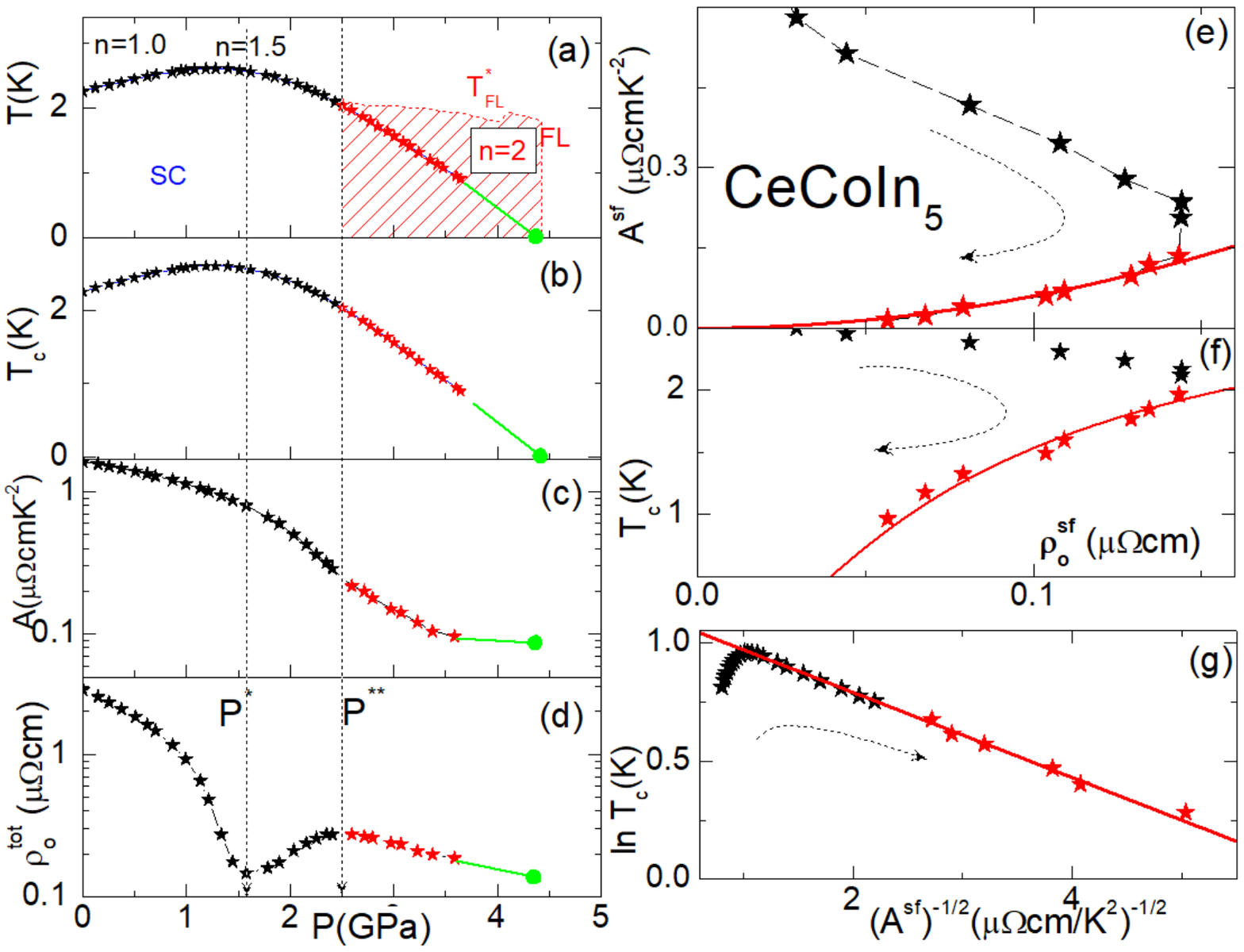}%
\caption{
\textbf{(a)} \textit{P–T} phase diagram and baric evolution of \textbf{(b)} $T_c(P)$, with a maximum at $P^{\ast}\!\approx\!1.6$\,GPa, and \textbf{(c)} $A(P)$ of \ce{CeCoIn5}.
\textbf{(d)} The baric evolution of $\rho^{\mbox{\tiny tot}}_{\mbox{\tiny 0}}(P)$, shown on a semilogarithmic scale, reveals a weak local maximum at $P^{\ast\ast}\!\approx\!2.5$\,GPa.
All datasets  were taken from the continuous lines (visual guides) of Ref.\,\protect\Onlinecite{Sidorov02-CeCoIn5-Q-Criticality}. Red symbols denote the QCF-driven FL regime; dashed lines are visual guides.
\textbf{(e)}  Correlation between $A^{\mbox{\tiny sf}}$ and $\rho^{\mbox{\tiny sf}}_{\mbox{\tiny 0}}(P)$. The red curve shows the fit to Eq.\ref{Eq.A-vs-rho} with $A_1\!\approx\!0$; the resulting parameters $A_0$ and $A_2$ are listed in Table~\ref{Tab.Fit-Values-A-Ro-Theta-F}.
\textbf{(f)}  $T_c$ \textit{versus} $\rho^{\mbox{\tiny sf}}_{\mbox{\tiny 0}}$. The red curve is a fit to Eq.\ref{Eq.lnTc-vs-ro-one}a.
\textbf{(g)}  $\ln T_c$ \textit{versus} $A^{-1/2}$. The solid red line is a fit to the linearized Eq.\ref{Eq.Tc-of-A}b; the fit parameters $\theta$ and $\mathcal{F}$ are listed in Table\,\ref{Tab.Fit-Values-A-Ro-Theta-F}.
Notably, the baric evolutions of $A$, $\rho^{\mbox{\tiny sf}}_{\mbox{\tiny 0}}$, and $T_c$ differ in the vicinity of $P^{\ast}$ and $P^{\ast\ast}$; nevertheless, above $P\!>\!P^{\ast\ast}$ these quantities exhibit correlated behavior. All fits are therefore confined to this QCF-driven FL range.
 }\label{Fig4-CeCoIn5}
\end{figure*}
%%%%%%%%%%%%%%%%%%    End Fig 4   %%%%%%%%%%%%%%%%%%%%%%%%%%%

%%%%%%%%%%%%%%%%%%%%%%  Begin Table 1     %%%%%%%%%%%%%%%%%%%%%%%%%%%%%%%%
%\begin{tabular}[l]{cm{0.25cm}cm{0.75cm}cm{0.75cm}cm{0.75cm}cm{0.5cm}cm{1.0cm}cm{1.0cm}cm{2.0cm}}
\begin{table*}
\scriptsize
\begin{center}
\caption{Representative fit parameters of \ce{CeCu2Ge2}, \ce{CeCu2Si2} and CeCoIn$_{5}$. Values of A$_o$ and A$_2$ were obtained from a fit of Eq.\ref{Eq.A-vs-rho}  to each of the $A - \rho^{\mbox{\tiny{sf}}}_{\mbox{\tiny{0}}}$ curves (Figs.\ref{Fig2-CeCu2Ge2}(e),\ref{Fig3-CeCu2Si2}(e),\ref{Fig4-CeCoIn5}(e)). As we found $A_1 \approx 0$, we fix $A_1 \equiv 0$. The values of  $\theta$ and $\mathcal{F}$ were obtained from a fit of Eq.\ref{Eq.Tc-of-A}b to Figs.\ref{Fig2-CeCu2Ge2}(g),\ref{Fig3-CeCu2Si2}(g),\ref{Fig4-CeCoIn5}(g). The low value of $\theta$ is a measure of the energy scale parameter.  It is worth noting that the strength and evolution of $\theta$ and $\mathcal{F}$ depend critically on the criteria for determining \textit{T$_{c}$} (10-90\%, 50\%, onset or zero-point etc.) and $A$  (the sample geometry, which is often not precisely determined). Additionally, the widely-different methods for sample synthesis and thermal treatment give rise to a corresponding variation in sample-dependent properties (e.g. non-stoichiometry, defects, impurities or disorder). Consequently, we expect a wide scatter in the fit parameters of different samples of the very same QCHF superconductor. It is reassuring that $\theta$ (and similarly $\mathcal{F}$) takes comparable values across different QCHF superconductors.  Finally, the fit range is confined to the boundaries of the QCF-driven FL regime ($P_L \le P \le P_H$, see text).}
\begin{tabular}[c]{ccccccc}
\hline\hline
$HFs$ & A$_o$ & A$_2$ & $ \theta$ & $\mathcal{F}$ & Fitting range  \\
- & $\mu \Omega \text{cm/K}^{2}$ & $ \times 10^6 \Omega^{-1} \text{cm}^{-1}/\text{K}^{2}$& K & $\times 10^{-3} (\Omega \text{cm/K}^{2})^{1/2}$& \\ %
\hline 
\ce{CeCu2Ge2}  & 0.00  & 7.2(2)$\times 10^{-5}$ & 5.1(3) & 0.43(2) & $P>P_{v}$  \\ %
\ce{CeCu2Si2}  & 0.01  & 4.0(2)$\times 10^{-5}$ & 2.5(2) & 0.49(4)& $P>P_{v}$  \\ %
%\ce{CeCoIn5}      & 0.00  & 2.8(2) & 4.2(2) & 0.39(2)& $P>P^{*}$ \\ %
\ce{CeCoIn5}      & 0.00  & 2.8(2) & 6.0(2) & 0.52(2)& $P>P^{**}$  \\ %
\hline
\label{Tab.Fit-Values-A-Ro-Theta-F}%
\end{tabular}
\end{center}
\end{table*} 
%%%%%%%%%%%%%%%%%%%%%%  End Table 1     %%%%%%%%%%%%%%%%%%%%%%%%%%%%%%%%
%
%%%%%%%%%%%%%%%%%% Begin Fig  5  %%%%%%%%%%%%%%%%%%%%%%%%%%%
\begin{figure*}[hbtp]
\centering
\includegraphics[scale=0.12,trim={0cm 0cm 0cm 0cm}, clip]{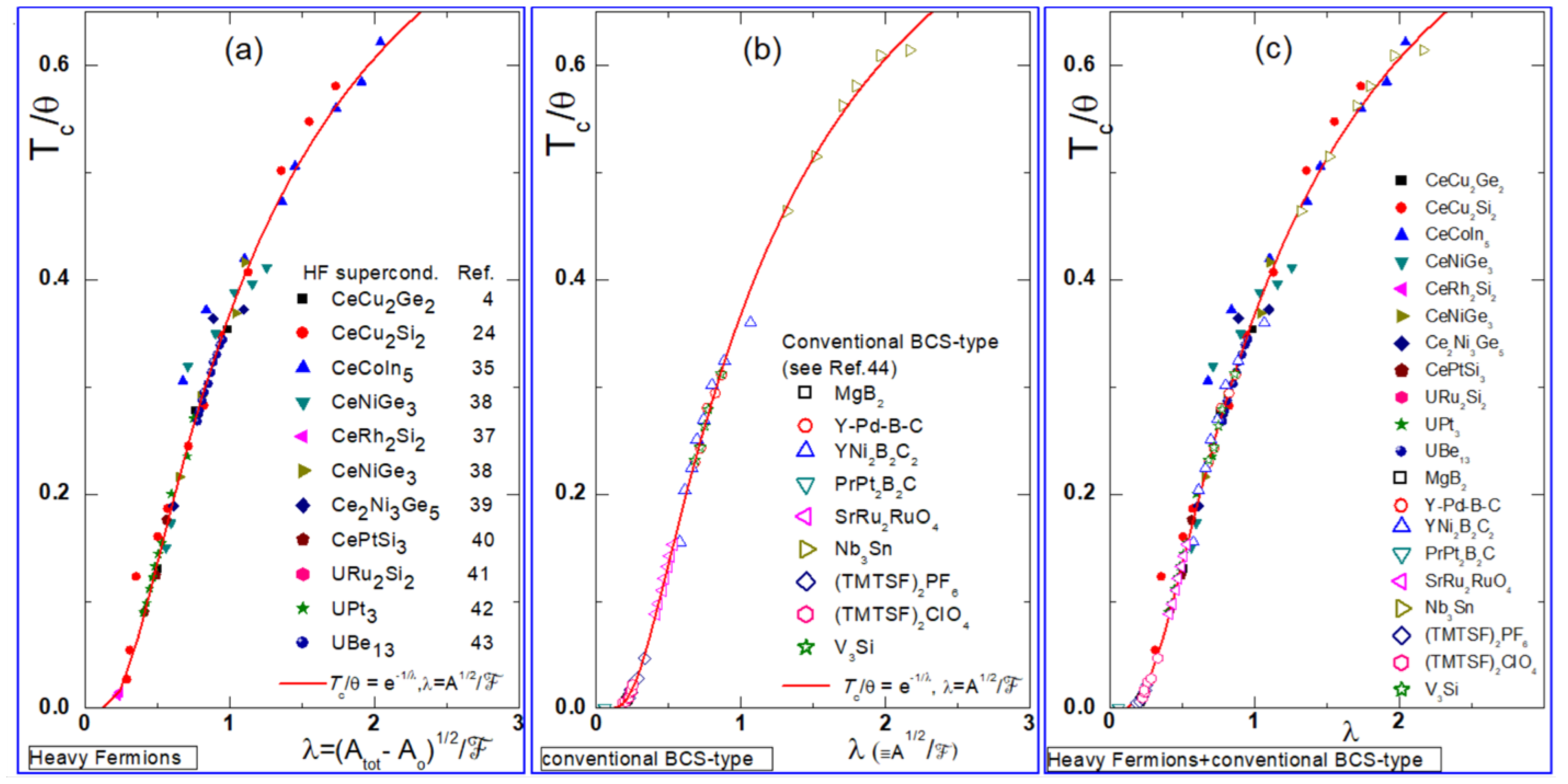}%
\caption{
\textbf{(a)}  Generalized plot of $\frac{T_{c}(P)}{\theta}$\textit{versus} $\lambda_{\text{eff}}=\frac{A_{\text{eff}}^{1/2}}{\mathcal{F}}$ of the tabulated QCHF superconductors. \cite{Jaccard99-HF-SC-Mag-HP, Holmes04-CeCu2Si2-ValenceFluctuation,Sheikin00-CeCu2Si2-HF-R-P,Sidorov02-CeCoIn5-Q-Criticality, Kotegawa06-CeNiG3-SC-AFM,Araki02-CeRh2Si2-SC-pressure,Nakashima04-CeNiGe3,Nakashima06-Ce2Ni3Ge5-SC-Pressure,Onuki08-CePt3Si-CeIr3Si,Hassinger08-URu2Si2-HiddenOrder-Nesting,deVisser84-UPt3-FL,
Ott83-UBe13-Unconv-SC}
\textbf{(b-c)}  For comparison, we include the plot of the same relation in conventional superconductors.\cite{Nunes12-FermiLiquid-SUC}  Similar plots were reported for pnictides [\protect\Onlinecite{18-Castro-Tc-A-Correlation}] and chalcogenides [\protect\Onlinecite{18-Soares-KxFe2-ySe2-Quantum-Conductance-Phase-Diagram}].
Each pair of sample-dependent  $\theta$ and $\mathcal{F}$ was evaluated within the pressure range wherein Eq.\ref{Eq.Tc-of-A} is valid [see Figs.\ref{Fig2-CeCu2Ge2}(g), \ref{Fig3-CeCu2Si2}(g), and \ref{Fig4-CeCoIn5}(g)]. Representative values of  $\theta$ and $\mathcal{F}$ are shown in Table\,\ref{Tab.Fit-Values-A-Ro-Theta-F}. %
It is noted that the pairs ($\frac{T_{c}(P)}{\theta}$,$\frac{A_{\text{eff}}^{1/2}}{\mathcal{F}}$) of QCHF superconductors are distributed within the same region occupied by conventional low-$T$ superconductors (see Fig.4 of
Ref.\,{\protect\Onlinecite{Nunes12-FermiLiquid-SUC}}): $\lambda_{\text{eff}}$ of each of QCHF superconductor is much below 2.2 (the upper limit for strong conventional superconductors\cite{Gurvitch86-Disorder-induced-transition-AT2}).
}%
\label{Fig4-Tc-A-AllHFS}
\end{figure*}
%%%%%%%%%%%%%%%%%%%  End Fig 5   %%%%%%%%%%%%%%%%%%%%%%%%%%%
\subsection{\texorpdfstring{CeCoIn$_{5}$ }{} \label{SubSec.CeCoIn5}}
The \textit{P–T} phase diagram of \ce{CeCoIn5} [Refs.\,\Onlinecite{Sidorov02-CeCoIn5-Q-Criticality,Ronning06-CeCoIn5-Pressure-QCP,Paglione03-CeCoIn5-H-QCP,Bianchi03-CeCoIn5-H-QCP}] is shown in Fig.\ref{Fig4-CeCoIn5}(a), while the pressure evolution of $T_c(P)$, $A^{\mbox{\tiny tot}}(P)$, and $\rho^{\mbox{\tiny tot}}_{\mbox{\tiny 0}}(P)$ appears in Figs.\ref{Fig4-CeCoIn5}(b–d).
Comparing the key features of \ce{CeCoIn5} with those of \ce{CeCu2X2} ($X=$ Si, Ge) reveals broad similarities within the QCF-driven FL regime, in particular:
\begin{itemize}
    \item $A^{\mbox{\tiny sf}}$ is large yet remains \textit{quadratic in} $\rho^{\mbox{\tiny sf}}_{\mbox{\tiny 0}}$ [Fig.\ref{Fig4-CeCoIn5}(e)];
    \item $T_c$ depends \textit{exponentially} on $1/\rho^{\mbox{\tiny sf}}_{\mbox{\tiny 0}}$ [Fig.\ref{Fig4-CeCoIn5}(f)];
    \item $\ln(T_c)$ is \textit{linear} in $\big(A^{\mbox{\tiny sf}}\big)^{-1/2}$ [Fig.\ref{Fig4-CeCoIn5}(g)].
\end{itemize}

It is worth noting that, despite similarities with \ce{CeCu2X2}, \ce{CeCoIn5} exhibits two distinct differences.\cite{Sidorov02-CeCoIn5-Q-Criticality,Ronning06-CeCoIn5-Pressure-QCP,Paglione03-CeCoIn5-H-QCP,Bianchi03-CeCoIn5-H-QCP}
(i) Its magnetic QCP lies at \emph{negative} pressure (on the extrapolated side of the pressure axis).
(ii) The maximum in $T_c$, occurring at $P^{\ast}\!\approx\!1.6$\,GPa, is clearly separated from the maximum in $\rho^{\mbox{\tiny sf}}_{\mbox{\tiny 0}}$, which appears at $P^{\ast\ast}\!\approx\!2.5$\,GPa.

We now summarize the experimental findings.
Despite substantial differences in intrinsic material properties, the following observations hold across the examined compounds:
\begin{enumerate}
\item There is a clear similarity in the overall evolution of the \textit{P–T} phase diagrams of QCHF superconductors, including the pressure dependences of $T_c(P)$, $A^{\mbox{\tiny sf}}(P)$, and $\rho^{\mbox{\tiny sf}}_{\mbox{\tiny 0}}(P)$.
\item Both the QCF-driven FL and superconducting phases exhibit nontrivial characteristics that depart markedly from conventional behavior (see Appendix~\ref{Sec.A-Comparison-FLs}).
\item Three robust correlations are observed across all systems (see Figs.\ref{Fig2-CeCu2Ge2}, \ref{Fig3-CeCu2Si2}, \ref{Fig4-CeCoIn5}). In particular, Fig.\ref{Fig4-Tc-A-AllHFS} highlights a universal scaling among all examined QCHF superconductors:
$\frac{T_c(P)}{\theta} \;\text{vs.}\; \lambda_{\mathrm{eff}}
\quad\text{with}\quad
\lambda_{\mathrm{eff}} \equiv \frac{A_{\mathrm{eff}}^{1/2}}{\mathcal{F}},$
where $\theta$ is the characteristic spin-fluctuation energy scale, $A_{\mathrm{eff}}$ is the fluctuation-enhanced $T^2$ coefficient after background subtraction, and $\mathcal{F}$ is the momentum-relaxation efficiency factor (see \S\ref{SubSec.Tc-A-Correlation}). This scaling underscores a shared underlying mechanism linking superconductivity and inelastic scattering across \emph{all} systems shown in Fig.\ref{Fig4-Tc-A-AllHFS}.
\end{enumerate}
%
%%%%%%%%%%%%%%%%%%%%%%%%%%%%%%%% Section %%%%%%%%%%%%%%%%%%%%%%%%%%%%%%%%%%%%%%%
\section{Theoretical analysis $-$ correlating \protect{\texorpdfstring{$\rho^{\mbox{\tiny{sf}}}_{\mbox{\tiny{0}}}$, $A$, and \textit{T$ _{c}$} }{} } \label{Sec.Theoretical-Derivation}}
The empirical analysis of the previous section yields two principal results: (i) a fluctuation-induced residual resistivity, $\rho^{\mbox{\tiny sf}}_{\mbox{\tiny 0}}$, that is strongly enhanced in the vicinity of the QCP and decreases upon moving away from it; and (ii) three correlations:
$\ln (\nicefrac{T_c}{\theta}) \propto A^{-1/2},\quad A^{\mbox{\tiny{sf}}}\propto \big(\rho^{\mbox{\tiny sf}}_{\mbox{\tiny 0}}\big)^2,\quad
\ln (\nicefrac{T_c}{\theta}) \propto \big(\rho^{\mbox{\tiny sf}}_{\mbox{\tiny 0}}\big)^{-1}$.
These findings imply the threefold role for QCFs schematized in Fig.\ref{Fig1-Spin_Fluctuations_Rho0_Tc_A}.
Below we show that, starting from Eqs.\ref{Eq-Kondo-Lattice-Hamiltonian}–\ref{Eq.Effective-Interaction-Potential}, one can compute  $\Gamma(T,\omega)$ and $\lambda(T,\omega)$ in terms of $V^{\mathrm{sf}}_{ee}(\mathbf q,\omega)$ and $\chi_m(\mathbf q,\omega)$. From these, we derive closed-form expressions for $\rho^{\mbox{\tiny sf}}_{\mbox{\tiny 0}}$, $T_c(\ell)$, and  $A^{sf}(\ell)$ \textemdash{} and their correlations \textemdash{} in terms of a characteristic fluctuation length $\ell$ (which tracks $\rho^{\mbox{\tiny sf}}_{\mbox{\tiny 0}} \sim \ell^{-1}$). The resulting relations reproduce the observed pressure trends and provide a transparent interpretation of why $T_c$ grows with stronger inelastic scattering (larger $A^{sf}$) while being exponentially sensitive to the fluctuation-controlled elastic channel encoded in $\rho^{\mbox{\tiny sf}}_{\mbox{\tiny 0}}$.
%
%%%%%%%%%%%%%%%%%%%%%%%%%%%%% Section %%%%%%%%%%%%%%%%% {\textstyle\mathstrut}
\subsection{Fluctuation induced residual resistivity $\rho^{\mbox{\tiny{sf}}}_{\mbox{\tiny{0}}}$
\label{SubSec.QC-Fluctuation-Role}}
To compute the scattering rate of conduction electrons off  AF spin fluctuations, we start from the general expression for the transition rate derived from Fermi's Golden Rule. The scattering rate for a conduction electron with momentum $\mathbf{k}$ is given by summing over all possible final states $\mathbf{k}'$ and energy transfers $\omega$:
\begin{eqnarray}
\Gamma=\frac{1}{\tau} &=& \frac{2\pi \mathbf{I}^2}{\hbar} \sum_{\mathbf{k}'} \int_{-\infty}^\infty d\omega \, \left| \langle \mathbf{k}' | \mathbf{s}_c(\mathbf{q}) | \mathbf{k} \rangle \right|^2 \text{Im} \chi(\mathbf{q}, \omega) \nonumber\\
& &\times \left[ n_B(\omega) + f(\epsilon_{\mathbf{k}'}) \right] \delta(\epsilon_{\mathbf{k}} - \epsilon_{\mathbf{k}'} - \omega), \label{Eq.Scattering-Rate}
\end{eqnarray}
where $\mathbf{q} = \mathbf{k} - \mathbf{k}'$, $n_B(\omega)$ is the Bose-Einstein distribution, and $f(\epsilon_\mathbf{k})$ is the Fermi-Dirac function. Here, $\text{Im} \chi(\mathbf{q}, \omega)$ encodes the spectrum of spin fluctuations.

The transport treatment used below is restricted to the experimentally identified QCF-driven FL regime on the paramagnetic side of the QCP, where the resistivity exponent has returned to $n=2$ and well-defined quasiparticles are restored. Accordingly, the formulas derived in this section should be understood as effective scaling relations for that restored-quasiparticle FL window, rather than as an asymptotically exact description of the immediate critical regime.

To understand the origin of a fluctuation-induced \emph{residual} resistivity ($T \rightarrow 0$), we focus on the elastic scattering channel provided by the zero-frequency spectral weight of these fluctuations. We model the spin susceptibility near the AF QCP using the Ornstein-Zernike (OZ) form:
\begin{equation}
\chi(\mathbf{q}, \omega, \delta) = \frac{\chi_0 \xi^2(\delta)}{1 + \xi^2(\delta) (\mathbf{q} - \mathbf{Q})^2 - i \omega / \Gamma_{sf}(\delta)}.
\label{Ornstein-Zernike}
\end{equation}
The correlation length $\xi(\delta) = \xi_0 |\delta|^{-\nu}$ diverges at the QCP ($\delta \to 0$), where $\delta = (P - P_c)/P_c$ is the dimensionless parameter controlling the proximity to criticality.\cite{Sachdev99-Quantum-Phase-Transition} The damping rate $\Gamma_{sf}(\delta) \sim \xi^{-z}(\delta)$, with $z = 2$ for Landau damping,\cite{Millis93-nonZero-Temperature-on-QCP-Itinerat-Fermi} ensures that spin fluctuations become increasingly long-lived as $\delta \to 0$.

The dominant contribution to the \emph{residual} scattering rate arises from the static $(\omega \to 0)$ component of the susceptibility (recall that $U(\mathbf{q}) = \mathbf{I}^2 \chi(\mathbf{q}, \omega=0, \delta)$, see Eq.\ref{Eq.Effective-Interaction-Potential}). For a plane wave state $\mathbf{k}$, the scattering rate $1/\tau_{\mathbf{k}}$ scales as the integral over all final states $\mathbf{k}'$ of $|U(\mathbf{q})|^2$, where $\mathbf{q} = \mathbf{k} - \mathbf{k}'$. Within the scaling-level transport approximation appropriate to the selected long-$\ell$ FL window, and assuming an effective spherical average over the relevant scattering processes, the transport scattering rate scales as
\begin{equation}
\frac{1}{\tau(\delta)} \propto \int d^3q \, |U(\mathbf{q})|^2.
\end{equation}
Substituting the static OZ form, $\chi(\mathbf{q}, 0, \delta) = \chi_0 \xi^2 / [1 + \xi^2 (\mathbf{q} - \mathbf{Q})^2]$, and changing variables to $\mathbf{p} = \xi (\mathbf{q} - \mathbf{Q})$, the integral becomes
\begin{equation}
\frac{1}{\tau(\delta)} \propto \frac{\mathbf{I}^4 \chi_0^2 \xi^4}{\xi^3} \int \frac{d^3p}{(1 + p^2)^2} \propto \mathbf{I}^4 \chi_0^2 \, \xi(\delta).
\end{equation}
The residual resistivity is given by the Drude formula, $\rho^{\mbox{\tiny{sf}}}_{\mbox{\tiny{0}}} = m^{\star}/(n e^2 \tau)$, where $m^{\star} = m_{\scriptscriptstyle o}(1+\lambda)$ is the heavy-Fermion effective mass, which is also renormalized near the QCP. Within a self-consistent spin-fluctuation picture, the quasiparticle mass is expected to increase upon approaching criticality from the FL side. For the present scaling analysis, we adopt the phenomenological crossover form $m^{\star}\propto \xi$, to capture the enhancement of renormalization produced by remnant quantum-critical fluctuations within the restored-quasiparticle FL regime. Combining these results yields the fundamental scaling for the fluctuation-induced residual resistivity:
\begin{equation}
\rho^{\mbox{\tiny sf}}_{\mbox{\tiny 0}}(\delta) \propto m^{\star} / \tau^{\mbox{\tiny{sf}}}_{\mbox{\tiny{0}}} \propto  \xi^2(\delta) \propto |\delta|^{-2\nu},
\label{eq:rho_scaling}
\end{equation}
 where  $\tau^{\mbox{\tiny{sf}}}_{\mbox{\tiny{0}}}$ denotes the characteristic spin-fluctuation scattering time associated with the distinct channel contributing to $\rho^{\mbox{\tiny{sf}}}_{\mbox{\tiny{0}}}$. 
This use of $m^{\star}\propto \xi$ should be understood as a \emph{crossover scaling input on the FL side}, not as an asymptotic critical statement at $\delta=0$. The more singular non-Fermi-liquid regime, in which the critical dynamics dominate transport and the FL description breaks down, lies outside the domain of applicability of the present treatment.
This divergent behavior is cut off at the QCP itself. A phenomenological form that captures this saturation and the decay on the $\delta > 0$ side is given by
\begin{equation}
\rho^{\mbox{\tiny sf}}_{\mbox{\tiny 0}}(\delta) = \frac{\rho_{\scriptscriptstyle max}}{1 + b_+ \delta^{2\nu}},
\label{eq:rho_paramagnetic}
\end{equation}
where $\rho_{\scriptscriptstyle max}$ is the peak resistivity at the QCP and $b_+$ is a nonuniversal constant. This form has the correct asymptotic scaling $\rho^{\mbox{\tiny{sf}}}_{\mbox{\tiny{0}}} \propto \delta^{-2\nu}$ for $\delta \gg 0$, following from Eq.\eqref{eq:rho_scaling}.

Below the QCP ($\delta < 0$), the OZ form must be modified to account for the (dimensionless) AFM gap $\Delta$ in the spin-wave spectrum. The susceptibility becomes
\begin{equation}
\chi_{{\textstyle\mathstrut} afm}(\mathbf{q}, \omega, \delta) \sim \frac{\xi^2(\delta)}{1 + \xi^2(\delta)(\mathbf{q} - \mathbf{Q})^2 + \Delta^2 - i\omega/\Gamma_{sf}(\delta)},
\end{equation}
where $\Delta \sim |\delta|^\beta$ suppresses low-energy fluctuations ($\beta$ is the order parameter exponent). The presence of the gap reduces the phase space for scattering, leading to a more rapid decay of $\rho^{\mbox{\tiny sf}}_{\mbox{\tiny 0}}$. This is captured by the phenomenological form
\begin{equation}
\rho^{\mbox{\tiny sf}}_{\mbox{\tiny 0}}(\delta) = \frac{\rho_{\scriptscriptstyle max}}{1 + b_- |\delta|^{2\nu} + \Delta}.
\label{eq:rho_AFM}
\end{equation}
The asymmetry between Eqs.~\eqref{eq:rho_paramagnetic} and \eqref{eq:rho_AFM} thus originates from the gapped versus gapless spectra of spin fluctuations on the two sides of the QCP.

The peak at $\delta = 0$ reflects the OZ susceptibility's dependence on $\xi(\delta)$: as $\delta \to 0$, $\chi(\mathbf{Q}, 0, \delta) \sim \xi^2(\delta)$ diverges, maximizing the scattering rate. The pressure dependence of $\rho^{\mbox{\tiny sf}}_{\mbox{\tiny 0}}$ provides a clear signature of quantum criticality. 
The peak structure in $\rho^{\mbox{\tiny sf}}_{\mbox{\tiny 0}}(P)$ near $P \!\approx\! P_c$ is evident in Figs.\ref{Fig2-CeCu2Ge2}(d), \ref{Fig3-CeCu2Si2}(d), and \ref{Fig4-CeCoIn5}(d), although its maximum does not coincide with the maximum of $T_c$ (most probably due to disorder).
Remarkably, the peak width is governed by the correlation-length exponent $\nu$, providing a practical route to extract critical exponents from transport by fitting the line shape (after subtraction of noncritical backgrounds and accounting for $b_{\pm}$ and $\Delta$) in the vicinity of $P_c$.
 Finally, deviations from scaling at finite $T$ reveal crossover effects between quantum and thermal critical regimes.

It is important to stress that the present treatment does not attempt to resolve microscopically the full momentum anisotropy of the critical scattering problem. Instead, the transport effects of the fluctuation-mediated interaction are encoded at the phenomenological level through the momentum-relaxation efficiency factor $F_\ell$ and the structure factor $f_\ell$, which account for the kinematic effectiveness of momentum relaxation and for the available phase space on the Fermi surface. In this sense, the analysis retains the transport consequences of momentum selectivity without claiming a fully isotropic, vertex-free microscopic result valid at all scales.
 
%%%%%%%%%%%%%%%%%%%%%%%%%%%%% Section %%%%%%%%%%%%%%%%%
\subsection{Scaling Length $\texorpdfstring{\ell}{}$ for Elastic Scattering from Quantum-Critical Fluctuations
 \label{SubSec.B-QC-Fluctuation-Role}}
We can account for the elastic scattering by introducing an effective fluctuation-controlled elastic-scattering length, $\ell$, which quantifies the real-space range over which quasi-static QC modes scatter quasiparticles coherently:
\begin{equation}
\ell \equiv v_F\,\tau^{\mbox{\tiny{sf}}}_{\mbox{\tiny{0}}} \text{ and } \rho^{\mbox{\tiny{sf}}}_{\mbox{\tiny{0}}}
%=\frac{m^{\star}}{n e^2\,\tau_{0}^{\mathrm{sf}}}
=\frac{m^{\star} v_F}{n e^2}\,\frac{1}{\ell}\equiv \frac{C_\rho}{\ell} \Longrightarrow  \ell \propto \big(\rho^{\mbox{\tiny{sf}}}_{\mbox{\tiny{0}}}\big)^{-1},
\label{Eq-Effective-Scale-Length}
\end{equation}
where $v_F$ is the Fermi velocity, $n$ is the carrier density, and  $C_\rho = m^{\star} v_F/(n e^2)$ is a slowly varying material-dependent constant.
The use of $\ell=v_F\tau^{\mbox{\tiny sf}}_{\mbox{\tiny 0}}$ instead of $\tau^{\mbox{\tiny sf}}_{\mbox{\tiny 0}}$ alone is deliberate. While $\tau^{\mbox{\tiny sf}}_{\mbox{\tiny 0}}$ characterizes the time scale of the fluctuation-induced residual channel, $\ell$ provides a more transparent real-space transport interpretation as an effective \emph{fluctuation-controlled elastic mean free path}. This choice is particularly useful here because it allows the three empirical correlations to be written in a compact one-parameter form, namely $\rho^{\mbox{\tiny sf}}_{\mbox{\tiny 0}}\propto \ell^{-1}$, $A^{\mbox{\tiny sf}}\propto \ell^{-2}$, and $\ln(T_c/\theta)\propto \ell$, thereby highlighting the common fluctuation scale that controls residual scattering, inelastic transport, and pairing.
This way $\ell$ is regarded as the \emph{QCF-induced transport mean free path} extracted from the elastic-like residual channel. 

Close to the QCP and for elastic scattering off slow QC modes we take, on general grounds,
$ \ell^{-1} \propto \xi^{\alpha}\quad \Longrightarrow \quad \rho^{\mbox{\tiny{sf}}}_{\mbox{\tiny{0}}} 
\propto \xi^{\alpha}\propto |\delta|^{-\alpha \nu},$
where the exponent $\alpha>0$ encapsulates how the density/strength of quasi-static scatterers grows with criticality. This scaling produces the above-mentioned peak-like enhancement of $\rho^{\mbox{\tiny{sf}}}_{\mbox{\tiny{0}}}(P)$ with the divergent behavior being cut off at the QCP [see Figs.\ref{Fig2-CeCu2Ge2}(d), \ref{Fig3-CeCu2Si2}(d), and \ref{Fig4-CeCoIn5}(d)].

Based on $\rho^{\mbox{\tiny{sf}}}_{\mbox{\tiny{0}}}\!\propto\!\ell^{-1}$ (Eq.\ref{Eq-Effective-Scale-Length}), the empirical relations
$ A^{\mbox{\tiny sf}} \propto \big(\rho^{\mbox{\tiny{sf}}}_{\mbox{\tiny{0}}}\big)^2$, $\ln (\nicefrac{T_c}{\theta}) \propto A^{-1/2}$, and 
$\ln (\nicefrac{T_c}{\theta}) \propto \big(\rho^{\mbox{\tiny{sf}}}_{\mbox{\tiny{0}}}\big)^{-1}$ translate into compact $\ell$-scalings:
$A^{\mbox{\tiny sf}} \propto \ell^{-2}$ and $\ln (\nicefrac{T_c}{\theta}) \propto \ell$.
Thus $\ell$ serves as a single-parameter measure of QC strength that simultaneously controls elastic residual scattering, FL inelastic scattering, and the pairing scale (Fig.\ref{Fig1-Spin_Fluctuations_Rho0_Tc_A}).

\subsection{Expression of \texorpdfstring{$T_c(\ell)$}{}  \label{SubSec.Tc-Calculation}} 

 For determining $T_c(\ell)$, explicitly parameterized by $\ell$, we take, as usual, the zero-gap limit, $\Delta_\ell\rightarrow 0$, of the imaginary-time ($i\omega_n=i\pi T(2n-1)$) in Eliashberg equations:
\begin{widetext}
\begin{eqnarray}
\Delta_\ell(i\omega_n)Z_\ell(i\omega_n)&=&
\pi T\sum_m\left[\lambda_\ell(i\omega_m-i\omega_n)-\mu^*(\omega_c)\theta(\omega_c-|\omega_m|)\right]
\frac{\Delta_\ell(i\omega_m)}{\sqrt{\omega_m^2+\Delta_\ell^2(i\omega_m)}},\nonumber\\
Z_\ell(i\omega_n)&=&1+\frac{\pi T}{\omega_n}\sum_m
\lambda_\ell(i\omega_m-i\omega_n)\frac{\omega_m}{\sqrt{\omega_m^2+\Delta_\ell^2(i\omega_m)}},
\end{eqnarray}
\end{widetext}
Here, the Coulomb pseudopotential is given by:
\begin{equation}
\mu^*(\omega_c)=\frac{\mu}{1+\mu\ln{\left(\frac{\epsilon_F}{\omega_c}\right)}},
\end{equation}
which is defined in terms of the bare repulsive Coulomb interaction, $\mu=N(\epsilon_F)V_C$, and a cutoff frequency $\omega_c$. The electron-boson coupling, $\lambda_\ell$, is:
\begin{eqnarray}
\lambda_\ell(i\omega_m-i\omega_n)&=&
2\int_0^\infty \frac{\omega\; \alpha^2{\cal{F}_\ell}(\omega)}{(\omega_m-\omega_n)^2+\omega^2}d\omega\nonumber\\
&\equiv& N(\epsilon_F)V^{sf}_{ee}(i\omega_m-i\omega_n;\ell),
\end{eqnarray}
where $V^{sf}_{ee}(i\omega_m-i\omega_n;\ell)$ is the retarded, attractive, boson-mediated electron-electron interaction, and $N(\epsilon_F)$ is the electronic density of states at the Fermi level.

Following Allen and Dynes \cite{Allen-Dynes75-Tc-S-C}, we adopt a two-square-well model for the coupling:
\begin{equation}
    \lambda_\ell(i\omega_m-i\omega_n)= \begin{cases}
        \lambda_\ell       & \mbox{for} \quad |\omega_m|,|\omega_n|\ll\omega_c,\\
        0              & otherwise
        \end{cases}
\end{equation}
The coupling strength is defined as:
\begin{equation}
\lambda_\ell=2\int_0^\infty \frac{\omega\; \alpha^2{\cal{F}_\ell}(\omega)}{\omega_{opt}^2+\omega^2}d\omega\equiv N(\epsilon_F)V^{sf}_{ee}(\ell),
\label{Eq-Lambda-Two-Square-Well}
\end{equation}
where $\omega_{opt}$ is an optimal frequency at which $\lambda_\ell$ is maximized. It's worth noting that the standard choice for the electron-boson coupling in the two-square-well approximation is $\lambda(i\omega_m-i\omega_n)=\lambda(0)=\lambda$, where the Matsubara summations are performed with the assumption that $\omega_m=\omega_n$ for all terms. This assumption of an instantaneous interaction is valid for weakly-coupled superconductors, which are characterized by a bosonic spectrum heavily weighted at high frequencies. In this case, the average boson frequency, $\langle\omega\rangle_{\alpha^2{\cal F}}$, is significantly larger than any potential difference $|\omega_m-\omega_n|=2\pi |m-n| k_B T/\hslash \ll\omega_c$.

However, when spectral weight shifts toward lower frequencies, retardation effects become important, and we must allow for $m\neq n$ in the Matsubara sums. One simple choice is to consider $m-n=1$, which leads to $\omega_{opt}=2\pi k_B T_c /\hslash$ at the transition temperature. A more comprehensive approach, based on a physical argument by Carbotte \cite{Carbotte90-SCs-Boson-Exchanged-Reivew}, suggests an optimal frequency:
Consider a harmonic oscillator representing transverse spin-fluctions with frequency $\omega$. Its polarization will be maximal near $\omega \approx \omega_{opt}$. If an electron with Fermi velocity $v_F$ passes by and excites the oscillator, such polarization will only affect a second electron (with the same velocity) within a region of size given by the coherence length, $\xi_0$. If the oscillation is too slow ($\omega \ll \omega_{opt}$), no polarization effects will be transmitted to the second electron. If the oscillation is too fast ($\omega \gg \omega_{opt}$), the polarization will average to zero before the second electron leaves the coherence perimeter. Therefore, the retarded interaction must vanish in both the $\omega\rightarrow 0$ and $\omega\rightarrow\infty$ limits, reaching a maximum at $\omega_{opt}$. This optimal frequency can be expressed as $\omega_{opt}=\pi v_F/2\xi_0$ in terms of $v_F$ and $\xi_0$.

Alternatively, an estimate for $\omega_{opt}$ that includes various Matsubara frequencies (satisfying $|\omega_m|,|\omega_n|\ll\omega_c$) is approximately $\omega_{opt}\approx 10 k_B T^{max}_c /\hslash$, which is slightly higher than the $m-n=1$ lower bound of $2\pi k_B T_c /\hslash$. The presence of a non-zero optimal frequency, $\omega_{opt}$, implies that the retarded, effective $V^{sf}_{ee}(\ell)$ is most significant around $\omega_{opt}$ and diminishes to zero at both the $\omega\rightarrow\infty$ and $\omega\rightarrow0$ limits.

Using this framework, we have $Z_\ell=1+\lambda_\ell$. The zero-gap limit, $\Delta_\ell\rightarrow 0$, of the Eliashberg equations then simplifies to:
\begin{equation}
1=\frac{\lambda_\ell-\mu^*}{1+\lambda_\ell}\pi T_c\sum_{|\omega_m|<\omega_c}
\frac{1}{|\omega_m|}\simeq \frac{\lambda_\ell-\mu^*}{1+\lambda_\ell}
\ln{\left[\frac{2e^\gamma \omega_c}{\pi T_c}\right]},
\end{equation}
where $\gamma\simeq 0.577$ is Euler's constant. Exponentiating both sides yields the well-known MacMillan \cite{Mcmillan68-Tc} or simplified Allen-Dynes \cite{Allen-Dynes75-Tc-S-C} expression for $T_{c}$:
\begin{equation}
T_c(\ell) = \theta \, \exp\left[ -\frac{1 + \lambda_\ell}{\lambda_\ell - \mu^*} \right], \mbox{ with }
\theta=\frac{1.13 \hbar\omega_c}{k_B }.
\label{Eq-Tc-McMillan}
\end{equation}
\subsection{Expression of \texorpdfstring{$A(\ell)$}{} \label{SubSec.A-Calculation} } 
On general grounds, the coefficient $A(\ell)$ is determined by the available phase space for momentum relaxation due to scattering. These are controlled by processes that modify the phase $\phi_{\mathbf q}$ through the electron–electron structure factor,
\begin{equation}
f_{\ell}(\mathbf k_1',\mathbf k_1,\mathbf k_2',\mathbf k_2)
=
\sum_{\mathbf q}
\overline{S}^{\,\ell}_{\mathbf q}(\mathbf k_1'-\mathbf k_1)\;
\overline{S}^{\,\ell}_{-\mathbf q}(\mathbf k_2'-\mathbf k_2),
\label{Eq-ee-Structure-Factor}
\end{equation}
where the quasi-momentum transfer is regulated by the convolution of two phase-weighted form factors $\overline{S}^{\,\ell}_{\mathbf q}$.
We compute $A(\ell)$ by considering electron–electron scattering that includes both the direct Coulomb interaction $V_C$ and the effective, retarded (boson-mediated) interaction $V^{\mathrm{sf}}_{ee}(\ell)$, with $\ell$ given in Eq.~\ref{Eq-Effective-Scale-Length}.

 We seek a variational solution, $\Phi_{{\bf k}}$, to the linearized Boltzmann transport equation, which allows the resistivity to be expressed as 
\begin{equation}
\rho_{ee}(T,\ell)=\frac{\langle\Phi_{{\bf k}},{\cal P}_\ell\Phi_{{\bf k}}\rangle}
{\left| \langle\Phi_{{\bf k}},X\rangle\right|^2}=A(\ell)T^2,
\end{equation}
where ${\cal P}_\ell$ is the scattering operator that maps the variational solution $\Phi_{{\bf k}}$ to a new momentum state ${\bf k}^\prime$, and $X\equiv X(E=1)$ represents the left side of the Boltzmann equation under a unit electric field. Integrating the expression, the normalization factor becomes:
\begin{equation}
\left|\langle\Phi_{{\bf k}},X\rangle\right|=\left| 2e\sum_{{\bf k}}v_{{\bf k}}\Phi_{{\bf k}}\left(-\frac{\partial f\left(\epsilon_{{\bf k}}\right)}{\partial\epsilon_{{\bf k}}}\right)\right|=\frac{e k_F^3}{3\pi^2\hbar m^{\star}}=\frac{n e}{m^{\star}},
\end{equation}
Here, $\Phi_{{\bf k}}=\vec{u}\cdot\vec{{\bf v}}_{{\bf k}}$ describes the deviation of the electron distribution from equilibrium, with $\vec{u}$ being the unit vector aligned with the applied electric field. The quasiparticle velocity is given by ${\bf v}_{\bf k}=\hbar {\bf k}/m^{\star}$
%, where $m^{\star}=1+\lambda$ is the heavy-Fermion effective mass and ${\bf k}$ is the momentum of the carriers. 
In terms of these quantities, the numerator is written as 
\begin{widetext}
\begin{eqnarray}
\langle\Phi_{{\bf k}},{\cal P}_\ell\Phi_{{\bf k}}\rangle  =  
\frac{1}{2k_{B}T} &\sum_{{\bf k}_1, {\bf k}_2, {\bf k}_1^\prime, {\bf k}_2^\prime}&
f_{\epsilon_{{\bf k}_{1}}}f_{\epsilon_{{\bf k}_{2}}}(1-f_{\epsilon_{{\bf k}_{1}^{\prime}}})(1-f_{\epsilon_{{\bf k}_{2}^{\prime}}})
\delta(\epsilon_{{\bf k}_1}+\epsilon_{{\bf k}_2}-\epsilon_{{\bf k}_1^\prime}-\epsilon_{{\bf k}_2^\prime})\nonumber\\
&\times&
(\Phi_{{\bf k}_{1}}+\Phi_{{\bf k}_{2}}-\Phi_{{\bf k}_{1}^{\prime}}-\Phi_{{\bf k}_{2}^{\prime}})^{2}
\Gamma_{{\bf k}_{1}+{\bf k}_{2}\rightarrow {\bf k}_{1}^{\prime}+{\bf k}_{2}^{\prime}}(\ell),
\label{Eq-Numeral-Boltzmann}
\end{eqnarray}
\end{widetext}
where $\Gamma_{{\bf k}_{1}+{\bf k}_{2}\rightarrow {\bf k}_{1}^{\prime}+{\bf k}_{2}^{\prime}}(\ell)$ is the transition amplitude for the total electron-electron interaction, $V_{tot}(\ell)$, and $f_{\epsilon_{{\bf k}}}$ is the equilibrium electron distribution:
\begin{equation}
f_{\epsilon_{{\bf k}}}=\frac{1}{e^{\beta(\epsilon_{{\bf k}}-\mu)}+1},
\end{equation}
with $\beta=1/k_{B}T$.
% and $\mu=\epsilon_{F}$, the Fermi energy.

It's important to note that as a system approaches the superconducting transition from the Fermi-liquid state ($T\rightarrow T_c^+(\ell)$), the total electron-electron interaction $V^0_{tot}(\ell)=V_{C}-V^{sf}_{ee}(\ell)$ is renormalized to:
\begin{equation}
V^0_{tot}(\ell)\rightarrow V_{tot}(\ell)=\frac{V_C-V^{sf}_{ee}(\ell)}{Z_\ell}=\frac{V_C-V^{sf}_{ee}(\ell)}{1+\lambda_\ell},
\label{Eq-Renormalized-Vtot(l)}
\end{equation}
The renormalization constant, $Z_\ell$, is identical to the one found when approaching from the superconducting ground state ($T\rightarrow T_c^-(\ell)$). This is a result that is asymptotically exact and was obtained using a renormalization group method that treats both the direct and effective components of the total interaction on equal footing.\cite{Tsai05-Renormalization-Strong-coupled-SUCs}

At low temperatures, we can constrain all electron states to the Fermi surface ($|{\bf k}_i|=k_F$). This allows us to convert the summation $\sum_{{\bf k}_i}$ into integrals over the electron energies $\epsilon_{{\bf k}_i}$ and solid angles, assuming a constant electronic density of states at the Fermi level. The energy conservation constraint can also be expressed in terms of the transferred energy to spin-fluctuations, $\hbar\omega$:
\begin{equation}  
\delta(\epsilon_{{\bf k}_1}+\epsilon_{{\bf k}_2}-\epsilon_{{\bf k}_1^\prime}-\epsilon_{{\bf k}_2^\prime})=
\int_{-\infty}^{\infty} d\omega \;   \delta(\epsilon_{{\bf k}_1^\prime}-\epsilon_{{\bf k}_1}-\hbar\omega) \delta(\epsilon_{{\bf k}_2^\prime}-\epsilon_{{\bf k}_2}+\hbar\omega),
\end{equation}  
This substitution enables the elimination of both $\epsilon_{{\bf k}_1^{\prime}}$ and $\epsilon_{{\bf k}_2^{\prime}}$, leaving us with:
\begin{eqnarray}
&{}&\int d\epsilon_{{\bf k}_{1}}\int d\epsilon_{{\bf k}_{2}}\:
f_{\epsilon_{{\bf k}_{1}}}f_{\epsilon_{{\bf k}_{2}}}
(1-f_{\epsilon_{{\bf k}_{1}}+\hbar\omega})(1-f_{\epsilon_{{\bf k}_{2}}-\hbar\omega})\nonumber\\
&=&\frac{\hbar^{2}\omega^{2}}{(e^{\beta\hbar\omega}-1)(1-e^{-\beta\hbar\omega})}.
\end{eqnarray}
Finally, integrating over the transferred energy $\hbar\omega$ yields:
\begin{equation}
\frac{1}{2 k_B T}\int_{-\infty}^{\infty}d(\hbar\omega)
\frac{\hbar^{2}\omega^{2}}{(e^{\beta\hbar\omega}-1)(1-e^{-\beta\hbar\omega})}=
\frac{\pi^{2}}{3}(k_{B}T)^{2},
\label{Eq-R-prop-T2}
\end{equation}
This result confirms that the resistivity contribution from this boson-mediated electron-electron interaction exhibits the characteristic Fermi-liquid quadratic-in-T dependence: $\rho_{ee}(T,\ell)=A(\ell) T^2$. The coefficient $A(\ell)$ is given by:
\begin{widetext}
\begin{equation}
A(\ell)=\left(\frac{m^{\star}}{ne}\right)^2\frac{\pi^2 k_B^2}{3(\hbar v_F)^4}
\int\int\int\int\frac{d\Omega_{{\bf k}_1}d\Omega_{{\bf k}_1^\prime}
d\Omega_{{\bf k}_2}d\Omega_{{\bf k}_2}^\prime}{(2\pi)^{12}}
(\Phi_{{\bf k}_{1}}+\Phi_{{\bf k}_{2}}-\Phi_{{\bf k}_{1}^{\prime}}-\Phi_{{\bf k}_{2}^{\prime}})^{2}\times
\Gamma_{{\bf k}_{1}+{\bf k}_{2}\rightarrow {\bf k}_{1}^{\prime}+{\bf k}_{2}^{\prime}}(\ell),
\label{Eq-A-N.Gamma-Vee^2}
\end{equation}
\end{widetext}
where $\Gamma_{{\bf k}_{1}+{\bf k}_{2}\rightarrow {\bf k}_{1}^{\prime}+{\bf k}_{2}^{\prime}}(\ell)$ is calculated using Fermi's golden rule:
\begin{equation}
\Gamma_{{\bf k}_{1}+{\bf k}_{2}\rightarrow {\bf k}_{1}^{\prime}+{\bf k}_{2}^{\prime}}(\ell)=
\left(\frac{2\pi}{\hbar}\right)\left|V_{tot}(\ell)\right|^2 f_\ell({\bf k}_1+{\bf k}_2-{\bf k}_1^\prime-{\bf k}_2^\prime).
\label{Eq-Gamma-Vee^2}
\end{equation}

The electron-electron structure factor $f_\ell({\bf k}_1+{\bf k}_2-{\bf k}_1^\prime-{\bf k}_2^\prime)$ in Eq.\ref{Eq-ee-Structure-Factor} enforces all kinematic constraints. By projecting all states ${\bf k}_i$ onto the roughened Fermi surface, we derive:
\begin{equation}
A(\ell)=F^2_\ell\left|V_{tot}(\ell)\right|^2,
\label{FL-Coeff-A}
\end{equation}
where $F_{\ell}$ \textemdash{} representing the efficiency of momentum relaxation and the availability of phase space for scattering \textemdash{} is defined as:\cite{21-FL-SC-Defectal-Reconciled}
\begin{widetext}
\begin{equation}
F^2_\ell=F^2_{\mbox{\tiny{0}}}\int\int\int\int\frac{d\Omega_{{\bf k}_1}\, d\Omega_{{\bf k}_1^\prime}\, d\Omega_{{\bf k}_2}\, d\Omega_{{\bf k}_2^\prime}}{(2\pi)^{12}}
(\Phi_{{\bf k}_{1}}+\Phi_{{\bf k}_{2}}-\Phi_{{\bf k}_{1}^{\prime}}-\Phi_{{\bf k}_{2}^{\prime}})^{2}\times
f_\ell({\bf k}_1+{\bf k}_2-{\bf k}_1^\prime-{\bf k}_2^\prime),
\label{Eq-Fell-Efficiency-Phase-Availability}
\end{equation}
\end{widetext}
with
\begin{equation}
F^2_{\mbox{\tiny{0}}}=(2\pi/\hbar)(m^{\star}/ne)^2(\pi^2 k_B^2/3 \hbar^4 v_F^4)(1/N^2(\epsilon_F)).
\label{Eq-F_o}
\end{equation}

A precise calculation of $F^2_\ell$ necessitates a microscopic analysis that accounts for all possible momentum-transferring relaxation channels. This is true only when:
\begin{equation}
[\vec{u}\cdot({\bf k}_{1}+{\bf k}_{2}-
{\bf k}_{1}^{\prime}-{\bf k}_{2}^{\prime})]^2\times f_\ell({\bf k}_1+
{\bf k}_2-{\bf k}_1^\prime-{\bf k}_2^\prime)
\neq 0.
\end{equation}
As seen from the above equation, the ultimate value of the Fermi-liquid coefficient, $A(\ell)$, depends critically on the available phase space for scattering, which is a function of the Fermi surface topology. 

Applying a variational approach to the linearized version of Boltzmann's transport equations within the relaxation time approximation (wherein the inverse scattering time is calculated by the use of Fermi's golden rule) we obtain\cite{21-FL-SC-Defectal-Reconciled} $A=A(\ell)T^2$ and
\begin{equation}
A(\ell) = F^2_{\ell} \left( \frac{\lambda_\ell - \mu^*}{1 + \lambda_\ell} \right)^2.
\label{Eq.A-vs-Lambda}
\end{equation}
Within the FL phase, $\ell$ is long, $1\ll k_F\ell_p <\infty$, and $\rho^{\mbox{\tiny{sf}}}_{\mbox{\tiny{0}}}$ is small ($k_F$ is the Fermi wave number); as such, Eq.\ref{Eq.A-vs-Lambda} can be expanded around $\lambda_\infty$ of the host matrix as
\begin{equation}
A(\ell) \simeq A_\infty + a_1 (\delta\lambda) + a_2 (\delta\lambda)^2 + \mathcal{O}((\delta\lambda)^3),
\label{Eq.A-vs-deltaLambda}
\end{equation}
wherein $A_\infty = \frac{|\lambda_\infty-\mu^*|^2}{1+\lambda_\infty}$ refer to the negligibly small kinematically-constrained contributions from the crystalline host matrix; the second and third term \big[containing $\delta\lambda= \lambda(\ell)- \lambda_\infty$ and the coefficients $a_1=2F^2_{\ell}(|\lambda_\infty-\mu^*|(1+\mu^*))/(1+\lambda_\infty)^3$ and
$a_2=F^2_{\ell}(1+\mu^*-2|\lambda_\infty-\mu^*|)(1+\mu^*)/(1+\lambda_\infty)^4$\big] denote contributions from all kinematically unconstrained relaxation processes after incorporating the fluctuation-mediated channel. 

The second-order polynomial expression of $A(\ell)$, Eq.\ref{Eq.A-vs-deltaLambda}, is reminiscent of the empirical \textit{quadratic-in-$\rho^{\mbox{\tiny{sf}}}_{\mbox{\tiny{0}}}$} of Eq.\ref{Eq.dominant-A-defect}. Below we look for an analytical expression of $A(\rho^{\mbox{\tiny{sf}}}_{\mbox{\tiny{0}}})$. Let us start by recalling the discussion in the Introduction that, for a typical  low-temperature Fermi liquid,  $\Gamma(\omega) \sim\omega^2$.
Then, on considering the relevant energy scale to be set by $k_B T$, one obtains the characteristic FL {\it quadratic-in-$T$} resistivity. Intuitively, this is related to the fact that, for a given temperature $T$, $N(E_F) k_B T$ single-particles within the Fermi surface are participating in the fluctuation-mediated two-particle channel (each single-particle can scatter into one another via this fluctuation-mediated scattering process): this generates the famous $AT^2$ resistivity contribution.

We expect $\rho^{\mbox{\tiny{sf}}}_{\mbox{\tiny{0}}} \propto (\delta\lambda)$ and that each is a function of $\ell^{-1}$. This allows us to establish a correlation among $\rho^{\mbox{\tiny{sf}}}_{\mbox{\tiny{0}}}$ and each of $T_c$ and $A$ (each is a function of $\ell$). 
Guided by these considerations, as well as the empirical relation of Eq.\ref{Eq.dominant-A-defect}, we consider 
\begin{equation}
A(\ell)\simeq A_{\mbox{\tiny{0}}} + A_1 (\rho^{\mbox{\tiny{sf}}}_{\mbox{\tiny{0}}}) + A_2 (\rho^{\mbox{\tiny{sf}}}_{\mbox{\tiny{0}}})^2 ,
\label{Eq.A-vs-rho}
\end{equation}
where $A_i(i=0,1,2$) are functions of $F^2_{\ell}, ~\mu^*, \text{ and } \lambda_\infty$. Eq.\ref{Eq.A-vs-rho} suggests three limiting contributions to $A(\ell)$: (i) A  non-QCF-related  contribution (dominant $A_{\mbox{\tiny{0}}} \neq 0$ when $A_1 \approx 0$ and  $A_2 \approx 0$), (ii) a {\it linear-in-$T$} NFL-related contribution\cite{Yuan22-Scaling-Strange-Metal-Cuprates} or a single-particle Koshino-Taylor-type, $A \propto \rho^{\mbox{\tiny{sf}}}_{\mbox{\tiny{0}}}$, contribution (dominant  $A_1 \neq0$  when $A_{\mbox{\tiny{0}}} \approx 0$ and  $A_2 \approx 0$), and (iii) a fluctuation-related contribution (dominant $A_2 \neq0$ when  $A_{\mbox{\tiny{0}}} \approx 0$ and $A_1 \approx0$). Figs.\ref{Fig2-CeCu2Ge2}(e), \ref{Fig3-CeCu2Si2}(e), \ref{Fig4-CeCoIn5}(e) belong to the third case.

\subsection{Correlation between \texorpdfstring{$T_c(\ell)$}{Tc(l)} and \texorpdfstring{$A(\ell)$}{A(l)} \label{SubSec.Tc-A-Correlation}}
We showed above that both $T_c(\ell)$ (Eq.\ref{Eq-Tc-McMillan}) and $A(\ell)$ (Eqs.\ref{Eq.A-vs-Lambda}) are governed by the same spin-fluctuation coupling $\lambda_\ell$. Eliminating $\lambda_\ell$ between these expressions yields the universal scaling relation, valid in the regime of large $\ell$ (small $\rho^{\text{\tiny{sf}}}_{\mbox{\tiny{0}}}$):
\begin{subequations}
\label{Eq.Tc-of-A}
\begin{align}
&T_c=\theta\,\exp\!\left[-\frac{\mathcal{F}}{\sqrt{A}}\right] \equiv \theta\,\exp\!\left[\frac{-1}{\lambda_{\text{eff}}}\right] \text{, or}\\
&\ln\!\left(\frac{T_c}{\theta}\right) \propto \left[\frac{1}{\sqrt{A}}\right].
\end{align}
\end{subequations}
For notational consistency with the empirical analysis, we set $\mathcal{F}\equiv F_{\ell}$, which is the same momentum-relaxation efficiency factor introduced in Eq.\ref{Eq-Fell-Efficiency-Phase-Availability}. These relations \textemdash{} which account for the experiments in Figs.\ref{Fig2-CeCu2Ge2}(g), \ref{Fig3-CeCu2Si2}(g), and \ref{Fig4-CeCoIn5}(g) \textemdash{} imply that, as pressure varies, the evolution of $T_c(\ell)$ with $A(\ell)$ is dictated solely by the kinematics of spin-fluctuation scattering: the system follows the universal curve given by Eq.\ref{Eq.Tc-of-A}a, as verified empirically in Fig.\ref{Fig4-Tc-A-AllHFS}(a).

Finally, in the regime where the quadratic dependence $A(\ell) \approx A_2\,[\rho_{\mbox{\tiny{0}}}^{\text{\tiny sf}}]^2$ dominates (Eq.\ref{Eq.A-vs-rho}), substituting into Eq.\ref{Eq.Tc-of-A} yields the approximate relation:
\begin{subequations}
\label{Eq.lnTc-vs-ro-one}
\begin{align}
&T_c \approx \theta \exp\!\left(-\frac{\mathcal{F}}{\sqrt{A_2} \, \rho_{\mbox{\tiny{0}}}^{\text{\tiny sf}}}\right) \text{, or }\\
&\ln\!\left(\frac{T_c}{\theta}\right) \propto \big(\rho_{\mbox{\tiny{0}}}^{\text{\tiny sf}}\big)^{-1},
\end{align}
\end{subequations}
which accounts for the empirical trends observed in Figs.\ref{Fig2-CeCu2Ge2}(f), \ref{Fig3-CeCu2Si2}(f), and \ref{Fig4-CeCoIn5}(f).

\section{Discussion and Conclusions  \label{Sec.Discussion-Conclusion}}
The similarity of the phase diagrams shown in Figs.\ref{Fig2-CeCu2Ge2}-\ref{Fig4-CeCoIn5}, as well as those of other QCHF superconductors,\cite{Gegenwart08-QC-HF-metals,Coleman07-HF-Review} suggests a generalized \textit{T-P} phase diagram that highlights the similarity in the cascade of distinct electronic states and in the overall evolution of $T_{c}(\ell)$, $A(\ell)$, and $\rho^{\mbox{\tiny{sf}}}_{\mbox{\tiny{0}}}(\ell)$. Of particular interest to this work is the nontrivial manifestation, in all QCHF superconductors, of a superconductivity, a FL character,  temperature-independent spin-fluctuation residual resistivity, $\rho^{\mbox{\tiny sf}}_{\mbox{\tiny 0}}(P)$, and the following  correlations:   $A_{eff} \propto (\rho^{\mbox{\tiny{sf}}}_{\mbox{\tiny{0}}})^2$,  $ln\frac{T_{c}}{\theta}\propto(\rho^{\mbox{\tiny{sf}}}_{\mbox{\tiny{0}}})^{\text{-1}}$ and $ln\frac{T_{c}}{\theta}\propto A_{eff}^{-\frac{1}{2}}$, with the latter being highlighted in Fig.\ref{Fig4-Tc-A-AllHFS} as a generalized plot of $\frac{T_{c}(\ell)}{\theta}$\textit{ versus } $\lambda_{\text{eff}}=\frac{A_{\text{eff}}^{1/2}}{\mathcal{F}}$. 

More broadly, the correlated FL regime emphasized in this work should not be viewed as a simple conventional background state. Rather, it is a fluctuation-shaped heavy-Fermi-liquid regime, in which the same control parameter that tunes the distance to a fluctuation-rich region also governs the renormalization quantities $\lambda$, $\gamma$, $m^{\star}$, and $A$. As the system approaches a QCP, a phase boundary, or a crossover regime from the FL side, the relevant fluctuations are enhanced and the corresponding quasiparticle renormalization grows; as the system is tuned away, these quantities decrease together. Within this framework, $m^{\star}$ is indeed one of the clearest indicators of the influence of quantum-critical fluctuations, but it is not an isolated one: it belongs to a broader and internally consistent set of heavy-fermion renormalization parameters that track the same underlying fluctuation physics.

The present transport treatment is intended only for the experimentally identified QCF-driven FL regime on the paramagnetic side of the QCP, where the resistivity exponent has already returned to $n=2$ and well-defined quasiparticles are restored. In that regime, the fluctuation spectrum remains strong enough to renormalize quasiparticles and leave measurable signatures in transport and pairing, yet sufficiently moderate that a semiclassical transport treatment remains meaningful. Accordingly, the identification of the quasiparticle scattering rate with the transport scattering rate is used here only as an effective approximation within this restored-quasiparticle FL window, not as an asymptotically exact description of the immediate critical regime. The same perspective applies to the Eliashberg and Boltzmann analyses that follow. In this sense, the factors $F_\ell$ and $f_\ell$ are introduced to encode, at the phenomenological level, the efficiency of momentum relaxation and the available phase space, without attempting a microscopic resolution of the full hot/cold anisotropy problem near the QCP.

Based on these arguments, the following inferences can be drawn: 
First, Figs.\ref{Fig2-CeCu2Ge2}–\ref{Fig4-CeCoIn5} show that, upon leaving the QCF-driven FL regime and approaching the QCP, $\rho^{\mbox{\tiny sf}}_{\mbox{\tiny 0}}$, $A$, and $T_c$ are simultaneously and continuously enhanced, with clear mutual correlations. This emphasizes that the growth of all three quantities is driven by an increase in a common interaction strength. However, the same figures also indicate that \emph{outside} the QCF-driven FL window our analytic expressions do not capture the full baric evolution of $\rho^{\mbox{\tiny sf}}_{\mbox{\tiny 0}}$, $A$, or $T_c$, a limitation attributable to the breakdown of the long-$\ell$ assumption. This shortcoming does not affect the validity of our analysis \emph{within} that FL regime.
It is worth emphasizing that the existence of an FL state is not a prerequisite for our approach; rather, the emergence of a dominant electron–electron scattering channel due to QCFs \emph{produces} the FL state, superconductivity, and their observed correlations.

Second, it has been reported that applying a magnetic field ($H>H_{2}$) within the superconducting dome ($P<P^{*}$) of CeCoIn$_{5}$ suppresses superconductivity and restores a FL normal state.\cite{Paglione03-CeCoIn5-H-QCP,Bianchi03-CeCoIn5-H-QCP,Bauer05-CeCoIn5-xSn-x-HCP} We attribute this field-induced FL character, emerging from the NFL regime, to a field-driven weakening of the spin fluctuations and, hence, to an increase in $\ell$.  This effect is analogous to that produced by increasing pressure in the QCHFSs discussed above, and is also consistent with the correlations observed in the $T$--$x$ and $T$--$B$ phase diagrams of the heavy-fermion system \ce{$Ce_xLa_{1-x}B_6$}.\cite{Jang17-Ce1-xLaxB6-Correlations-Effective-mass-Fluctuation,Nakamura06-Ce1-x-LaxB6-NFL-to-FL}

Third, we argued above that in the long-$\ell$ regime, spin fluctuations mediate the electron–electron interaction, which leads to the scaling between $T_c/\theta$ and $A_{\mathrm{eff}}$ (Fig.\ref{Fig4-Tc-A-AllHFS}a). 
The same scaling also appears in phonon-mediated superconductors, as illustrated for conventional systems in Fig.\ref{Fig4-Tc-A-AllHFS}b.\cite{21-FL-SC-Defectal-Reconciled,Nunes12-FermiLiquid-SUC} More relevant to the present work is the observation that analogous scaling has also been reported in Fe-based pnictides\cite{18-Castro-Tc-A-Correlation} and chalcogenides,\cite{18-Soares-KxFe2-ySe2-Quantum-Conductance-Phase-Diagram} suggesting that, as in heavy-fermion compounds, spin-fluctuation modes are the most plausible mediating bosons in these systems.

In summary, we investigated superconductivity, Fermi-liquid transport, and their correlations on the high-pressure side of quantum criticality (the QCF-driven FL regime) in representative QCHF superconductors. Empirically, upon tuning control parameters (pressure, alloying, defect incorporation): (i) the normal-state resistivity follows $\rho(T)=\rho^{\mbox{\tiny sf}}_{\mbox{\tiny 0}}+A T^2$ with $A\propto\big(\rho^{\mbox{\tiny sf}}_{\mbox{\tiny 0}}\big)^2$; (ii) the superconducting scale correlates with the residual channel as $\ln\!\big(T_c/\theta\big)\propto\big(\rho^{\mbox{\tiny sf}}_{\mbox{\tiny 0}}\big)^{-1}$; and (iii) $\ln\!\big(T_c/\theta\big)\propto A^{-1/2}$. These relationships are consistently observed across distinct compounds and collapse onto a universal $T_c$–$A$ scaling curve in this FL window.
We attribute these features—and their mutual correlations—to the threefold role of QCFs in this regime: mediating an effective pairing interaction, generating an inelastic quasiparticle scattering, and contributing an effective elastic residual scattering. The simultaneous emergence of these effects, together with their empirical interrelations, points to a unified fluctuation mechanism governing pairing, FL transport, and residual scattering.
Theoretically, within a Kondo-lattice–based description, we employed Migdal–Eliashberg theory and Boltzmann transport to derive analytic expressions for $\rho^{\mbox{\tiny sf}}_{\mbox{\tiny 0}}$, $A$, and $T_c$, and their interrelations. 
In the long-$\ell$ FL regime, our analyses reproduce the observed pressure trends and universal scalings, yielding a coherent framework that links the strength and kinematics of spin fluctuations to superconducting pairing, FL transport, and residual scattering in QCHF superconductors.

\begin{acknowledgments}
We acknowledge partial financial support through the INCT project Advanced Quantum Materials, involving the Brazilian agencies CNPq (Proc. 408766/2024-7), FAPESP (Proc. 2025/27091-3), and CAPES. M.B.S.N also acknowledges partial finantial support by the Brazilian funding agencies: CNPq through the grant 442072/2023-6, and FAPERJ through the grants E-26/210.100/2023 and E-26/210.781/2025. 
\end{acknowledgments}

\section*{Data availability}
The datasets used and/or analyzed during the current study available from the corresponding author on reasonable request.
%
%\section*{Funding}
%\begin{funding}
%\end{funding}
%\typeout{}
%\bibliographystyle{apsrev}
%\bibliography{ C:/OneDrive/RefBibTex/BIB-FermiLiquid-Conv-SC-Defectal, C:/OneDrive/RefBibTex/Bib-Chalcogenides, C:/OneDrive/RefBibTex/Bib-FermiLiquid-HeavyFermion-Kondo, C:/OneDrive/RefBibTex/Bib-pnictides, C:/OneDrive/RefBibTex/Bib-massalami, C:/OneDrive/RefBibTex/Bib-HTc-Cuprates-PseudoGap-QuantumPhases, C:/OneDrive/RefBibTex/Bib-HTc-Conventional-SUC,  C:/OneDrive/RefBibTex/Bib-ToBePublished-Comments,C:/OneDrive/RefBibTex/Bib-Thin-Films-Defectal}
%%
%\bibliography{RefBibTex/Bib-FermiLiquid-HeavyFermion-Kondo.bib,  RefBibTex/Bib-Chalcogenides.bib, RefBibTex/Bib-pnictides.bib, RefBibTex/Bib-massalami.bib, RefBibTex/Bib-HTc-Cuprates-PseudoGap-QuantumPhases.bib, RefBibTex/Bib-HTc-Conventional-SUC.bib, RefBibTex/Bib-ToBePublished-Comments,RefBibTex/Bib-Thin-Films-Defectal,RefBibTex/BIB-FermiLiquid-Conv-SC-Defectal}
%\bibliography{Tex-HF-SUC-Fl-2025-v5.bbl}

\begin{thebibliography}{62}%
\makeatletter
\providecommand \@ifxundefined [1]{%
 \@ifx{#1\undefined}
}%
\providecommand \@ifnum [1]{%
 \ifnum #1\expandafter \@firstoftwo
 \else \expandafter \@secondoftwo
 \fi
}%
\providecommand \@ifx [1]{%
 \ifx #1\expandafter \@firstoftwo
 \else \expandafter \@secondoftwo
 \fi
}%
\providecommand \natexlab [1]{#1}%
\providecommand \enquote  [1]{``#1''}%
\providecommand \bibnamefont  [1]{#1}%
\providecommand \bibfnamefont [1]{#1}%
\providecommand \citenamefont [1]{#1}%
\providecommand \href@noop [0]{\@secondoftwo}%
\providecommand \href [0]{\begingroup \@sanitize@url \@href}%
\providecommand \@href[1]{\@@startlink{#1}\@@href}%
\providecommand \@@href[1]{\endgroup#1\@@endlink}%
\providecommand \@sanitize@url [0]{\catcode `\\12\catcode `\$12\catcode
  `\&12\catcode `\#12\catcode `\^12\catcode `\_12\catcode `\%12\relax}%
\providecommand \@@startlink[1]{}%
\providecommand \@@endlink[0]{}%
\providecommand \url  [0]{\begingroup\@sanitize@url \@url }%
\providecommand \@url [1]{\endgroup\@href {#1}{\urlprefix }}%
\providecommand \urlprefix  [0]{URL }%
\providecommand \Eprint [0]{\href }%
\providecommand \doibase [0]{https://doi.org/}%
\providecommand \selectlanguage [0]{\@gobble}%
\providecommand \bibinfo  [0]{\@secondoftwo}%
\providecommand \bibfield  [0]{\@secondoftwo}%
\providecommand \translation [1]{[#1]}%
\providecommand \BibitemOpen [0]{}%
\providecommand \bibitemStop [0]{}%
\providecommand \bibitemNoStop [0]{.\EOS\space}%
\providecommand \EOS [0]{\spacefactor3000\relax}%
\providecommand \BibitemShut  [1]{\csname bibitem#1\endcsname}%
\let\auto@bib@innerbib\@empty
%</preamble>
\bibitem [{\citenamefont {Gegenwart}\ \emph {et~al.}(2008)\citenamefont
  {Gegenwart}, \citenamefont {Si},\ and\ \citenamefont
  {Steglich}}]{Gegenwart08-QC-HF-metals}%
  \BibitemOpen
  \bibfield  {author} {\bibinfo {author} {\bibfnamefont {P.}~\bibnamefont
  {Gegenwart}}, \bibinfo {author} {\bibfnamefont {Q.}~\bibnamefont {Si}},\ and\
  \bibinfo {author} {\bibfnamefont {F.}~\bibnamefont {Steglich}},\ }\bibfield
  {title} {\bibinfo {title} {Quantum criticality in heavy-fermion metals},\
  }\href@noop {} {\bibfield  {journal} {\bibinfo  {journal} {Nat. Phys.}\
  }\textbf {\bibinfo {volume} {4}},\ \bibinfo {pages} {186} (\bibinfo {year}
  {2008})}\BibitemShut {NoStop}%
\bibitem [{\citenamefont {Coleman}(2007)}]{Coleman07-HF-Review}%
  \BibitemOpen
  \bibfield  {author} {\bibinfo {author} {\bibfnamefont {P.}~\bibnamefont
  {Coleman}},\ }\bibfield  {title} {\bibinfo {title} {Heavy fermions: electrons
  at the edge of magnetism},\ }\href@noop {} {\bibfield  {journal} {\bibinfo
  {journal} {Handbook of Magnetism and Advanced Magnetic Materials}\ }
  (\bibinfo {year} {2007})}\BibitemShut {NoStop}%
\bibitem [{\citenamefont {Stewart}(1984)}]{Stewart84-HeavyFermion-Review}%
  \BibitemOpen
  \bibfield  {author} {\bibinfo {author} {\bibfnamefont {G.~R.}\ \bibnamefont
  {Stewart}},\ }\bibfield  {title} {\bibinfo {title} {Heavy-fermion systems},\
  }\href@noop {} {\bibfield  {journal} {\bibinfo  {journal} {Rev. Mod. Phys.}\
  }\textbf {\bibinfo {volume} {56}},\ \bibinfo {pages} {755} (\bibinfo {year}
  {1984})}\BibitemShut {NoStop}%
\bibitem [{\citenamefont {Jaccard}\ \emph {et~al.}(1999)\citenamefont
  {Jaccard}, \citenamefont {Wilhelm}, \citenamefont {Alami-Yadri},\ and\
  \citenamefont {Vargoz}}]{Jaccard99-HF-SC-Mag-HP}%
  \BibitemOpen
  \bibfield  {author} {\bibinfo {author} {\bibfnamefont {D.}~\bibnamefont
  {Jaccard}}, \bibinfo {author} {\bibfnamefont {H.}~\bibnamefont {Wilhelm}},
  \bibinfo {author} {\bibfnamefont {K.}~\bibnamefont {Alami-Yadri}},\ and\
  \bibinfo {author} {\bibfnamefont {E.}~\bibnamefont {Vargoz}},\ }\bibfield
  {title} {\bibinfo {title} {Magnetism and superconductivity in heavy fermion
  compounds at high pressure},\ }\href@noop {} {\bibfield  {journal} {\bibinfo
  {journal} {Physica B}\ }\textbf {\bibinfo {volume} {259-261}},\ \bibinfo
  {pages} {1 } (\bibinfo {year} {1999})}\BibitemShut {NoStop}%
\bibitem [{\citenamefont {Yuan}\ \emph {et~al.}(2003)\citenamefont {Yuan},
  \citenamefont {Grosche}, \citenamefont {Deppe}, \citenamefont {Geibel},
  \citenamefont {Sparn},\ and\ \citenamefont
  {Steglich}}]{Yuan03-CeCu2Si2-SC-Phases-HF}%
  \BibitemOpen
  \bibfield  {author} {\bibinfo {author} {\bibfnamefont {H.~Q.}\ \bibnamefont
  {Yuan}}, \bibinfo {author} {\bibfnamefont {F.~M.}\ \bibnamefont {Grosche}},
  \bibinfo {author} {\bibfnamefont {M.}~\bibnamefont {Deppe}}, \bibinfo
  {author} {\bibfnamefont {C.}~\bibnamefont {Geibel}}, \bibinfo {author}
  {\bibfnamefont {G.}~\bibnamefont {Sparn}},\ and\ \bibinfo {author}
  {\bibfnamefont {F.}~\bibnamefont {Steglich}},\ }\bibfield  {title} {\bibinfo
  {title} {Observation of two distinct superconducting phases in
  \textrm{CeCu$_{2}$Si$_{2}$}},\ }\href@noop {} {\bibfield  {journal} {\bibinfo
   {journal} {Science}\ }\textbf {\bibinfo {volume} {302}},\ \bibinfo {pages}
  {2104} (\bibinfo {year} {2003})}\BibitemShut {NoStop}%
\bibitem [{\citenamefont {Yang}\ and\ \citenamefont
  {Pines}(2014)}]{Yang14-SC-Heavy-Electrons}%
  \BibitemOpen
  \bibfield  {author} {\bibinfo {author} {\bibfnamefont {Y.-f.}\ \bibnamefont
  {Yang}}\ and\ \bibinfo {author} {\bibfnamefont {D.}~\bibnamefont {Pines}},\
  }\bibfield  {title} {\bibinfo {title} {Emergence of superconductivity in
  heavy-electron materials},\ }\href@noop {} {\bibfield  {journal} {\bibinfo
  {journal} {PNAS}\ }\textbf {\bibinfo {volume} {111}},\ \bibinfo {pages}
  {18178} (\bibinfo {year} {2014})}\BibitemShut {NoStop}%
\bibitem [{\citenamefont
  {Norman}(2011)}]{Norman11-Unconventional-SUC-Challenge}%
  \BibitemOpen
  \bibfield  {author} {\bibinfo {author} {\bibfnamefont {M.~R.}\ \bibnamefont
  {Norman}},\ }\bibfield  {title} {\bibinfo {title} {The challenge of
  unconventional superconductivity},\ }\href@noop {} {\bibfield  {journal}
  {\bibinfo  {journal} {Science}\ }\textbf {\bibinfo {volume} {332}},\ \bibinfo
  {pages} {196} (\bibinfo {year} {2011})}\BibitemShut {NoStop}%
\bibitem [{\citenamefont
  {Scalapino}(2012)}]{Scalapino12-UnconSCs-CommonThread}%
  \BibitemOpen
  \bibfield  {author} {\bibinfo {author} {\bibfnamefont {D.~J.}\ \bibnamefont
  {Scalapino}},\ }\bibfield  {title} {\bibinfo {title} {A common thread: The
  pairing interaction for unconventional superconductors},\ }\href@noop {}
  {\bibfield  {journal} {\bibinfo  {journal} {Reviews of Modern Physics}\
  }\textbf {\bibinfo {volume} {84}},\ \bibinfo {pages} {1383} (\bibinfo {year}
  {2012})}\BibitemShut {NoStop}%
\bibitem [{\citenamefont {Pines}(2013)}]{Pines13-PhaseDiagramCuprates}%
  \BibitemOpen
  \bibfield  {author} {\bibinfo {author} {\bibfnamefont {D.}~\bibnamefont
  {Pines}},\ }\bibfield  {title} {\bibinfo {title} {Finding new
  superconductors: The spin-fluctuation gateway to high \textrm{T$_c$} and
  possible room temperature superconductivity},\ }\href
  {https://doi.org/dx.doi.org/10.1021/jp403088e} {\bibfield  {journal}
  {\bibinfo  {journal} {J. Phys. Chem. B}\ }\textbf {\bibinfo {volume} {117}},\
  \bibinfo {pages} {13145} (\bibinfo {year} {2013})}\BibitemShut {NoStop}%
\bibitem [{\citenamefont {Keimer}\ \emph {et~al.}(2015)\citenamefont {Keimer},
  \citenamefont {Kivelson}, \citenamefont {Norman}, \citenamefont {Uchida},\
  and\ \citenamefont {Zaanen}}]{Keimer15-Cuprate-Quantum-Matter-HTC-SUC}%
  \BibitemOpen
  \bibfield  {author} {\bibinfo {author} {\bibfnamefont {B.}~\bibnamefont
  {Keimer}}, \bibinfo {author} {\bibfnamefont {S.~A.}\ \bibnamefont
  {Kivelson}}, \bibinfo {author} {\bibfnamefont {M.~R.}\ \bibnamefont
  {Norman}}, \bibinfo {author} {\bibfnamefont {S.}~\bibnamefont {Uchida}},\
  and\ \bibinfo {author} {\bibfnamefont {J.}~\bibnamefont {Zaanen}},\
  }\bibfield  {title} {\bibinfo {title} {From quantum matter to
  high-temperature superconductivity in copper oxides},\ }\href@noop {}
  {\bibfield  {journal} {\bibinfo  {journal} {Nature}\ }\textbf {\bibinfo
  {volume} {518}},\ \bibinfo {pages} {179} (\bibinfo {year}
  {2015})}\BibitemShut {NoStop}%
\bibitem [{\citenamefont
  {Stewart}(2017)}]{Stewart17-Unconventional-SUC-Review}%
  \BibitemOpen
  \bibfield  {author} {\bibinfo {author} {\bibfnamefont {G.~R.}\ \bibnamefont
  {Stewart}},\ }\bibfield  {title} {\bibinfo {title} {Unconventional
  superconductivity},\ }\href@noop {} {\bibfield  {journal} {\bibinfo
  {journal} {Advances in Physics}\ }\textbf {\bibinfo {volume} {66}},\ \bibinfo
  {pages} {75} (\bibinfo {year} {2017})}\BibitemShut {NoStop}%
\bibitem [{\citenamefont {Yang}\ \emph {et~al.}(2017)\citenamefont {Yang},
  \citenamefont {Pines},\ and\ \citenamefont
  {Lonzarich}}]{Yang17-QC-Scaling-FLuctuation-Kondo-Lattice}%
  \BibitemOpen
  \bibfield  {author} {\bibinfo {author} {\bibfnamefont {Y.-f.}\ \bibnamefont
  {Yang}}, \bibinfo {author} {\bibfnamefont {D.}~\bibnamefont {Pines}},\ and\
  \bibinfo {author} {\bibfnamefont {G.}~\bibnamefont {Lonzarich}},\ }\bibfield
  {title} {\bibinfo {title} {Quantum critical scaling and fluctuations in kondo
  lattice materials},\ }\href@noop {} {\bibfield  {journal} {\bibinfo
  {journal} {PNAS}\ }\textbf {\bibinfo {volume} {114}},\ \bibinfo {pages}
  {6250} (\bibinfo {year} {2017})}\BibitemShut {NoStop}%
\bibitem [{\citenamefont {Mathur}\ \emph {et~al.}(1998)\citenamefont {Mathur},
  \citenamefont {Grosche}, \citenamefont {Julian}, \citenamefont {Walker},
  \citenamefont {Freye}, \citenamefont {Haselwimmer},\ and\ \citenamefont
  {Lonzarich}}]{Mathur98-HF-SC-Mag-Mediation}%
  \BibitemOpen
  \bibfield  {author} {\bibinfo {author} {\bibfnamefont {N.~D.}\ \bibnamefont
  {Mathur}}, \bibinfo {author} {\bibfnamefont {F.~M.}\ \bibnamefont {Grosche}},
  \bibinfo {author} {\bibfnamefont {S.~R.}\ \bibnamefont {Julian}}, \bibinfo
  {author} {\bibfnamefont {I.~R.}\ \bibnamefont {Walker}}, \bibinfo {author}
  {\bibfnamefont {D.~M.}\ \bibnamefont {Freye}}, \bibinfo {author}
  {\bibfnamefont {R.~K.~W.}\ \bibnamefont {Haselwimmer}},\ and\ \bibinfo
  {author} {\bibfnamefont {G.~G.}\ \bibnamefont {Lonzarich}},\ }\bibfield
  {title} {\bibinfo {title} {Magnetically mediated superconductivity in heavy
  fermion compounds},\ }\href@noop {} {\bibfield  {journal} {\bibinfo
  {journal} {Nature}\ }\textbf {\bibinfo {volume} {394}},\ \bibinfo {pages}
  {39} (\bibinfo {year} {1998})}\BibitemShut {NoStop}%
\bibitem [{\citenamefont {Monthoux}\ \emph {et~al.}(2007)\citenamefont
  {Monthoux}, \citenamefont {Pines},\ and\ \citenamefont
  {Lonzarich}}]{Monthoux07-SC-Without-Phonon}%
  \BibitemOpen
  \bibfield  {author} {\bibinfo {author} {\bibfnamefont {P.}~\bibnamefont
  {Monthoux}}, \bibinfo {author} {\bibfnamefont {D.}~\bibnamefont {Pines}},\
  and\ \bibinfo {author} {\bibfnamefont {G.}~\bibnamefont {Lonzarich}},\
  }\bibfield  {title} {\bibinfo {title} {Superconductivity without phonons},\
  }\href@noop {} {\bibfield  {journal} {\bibinfo  {journal} {Nature}\ }\textbf
  {\bibinfo {volume} {450}},\ \bibinfo {pages} {1177} (\bibinfo {year}
  {2007})}\BibitemShut {NoStop}%
\bibitem [{\citenamefont {Wakimoto}\ \emph {et~al.}(2004)\citenamefont
  {Wakimoto}, \citenamefont {Zhang}, \citenamefont {Yamada}, \citenamefont
  {Swainson}, \citenamefont {Kim},\ and\ \citenamefont
  {Birgeneau}}]{Wakimoto04-HTC-Overdoped-Low-Excitation-SUC-Correlations}%
  \BibitemOpen
  \bibfield  {author} {\bibinfo {author} {\bibfnamefont {S.}~\bibnamefont
  {Wakimoto}}, \bibinfo {author} {\bibfnamefont {H.}~\bibnamefont {Zhang}},
  \bibinfo {author} {\bibfnamefont {K.}~\bibnamefont {Yamada}}, \bibinfo
  {author} {\bibfnamefont {I.}~\bibnamefont {Swainson}}, \bibinfo {author}
  {\bibfnamefont {H.}~\bibnamefont {Kim}},\ and\ \bibinfo {author}
  {\bibfnamefont {R.}~\bibnamefont {Birgeneau}},\ }\bibfield  {title} {\bibinfo
  {title} {Direct relation between the low-energy spin excitations and
  superconductivity of overdoped high-{T$_c$} superconductors},\ }\href@noop {}
  {\bibfield  {journal} {\bibinfo  {journal} {Phys. Rev. Lett.}\ }\textbf
  {\bibinfo {volume} {92}},\ \bibinfo {pages} {217004} (\bibinfo {year}
  {2004})}\BibitemShut {NoStop}%
\bibitem [{\citenamefont
  {Taillefer}(2010)}]{Taillefer10-Scattering-pairing-HTc-Cuprate}%
  \BibitemOpen
  \bibfield  {author} {\bibinfo {author} {\bibfnamefont {L.}~\bibnamefont
  {Taillefer}},\ }\bibfield  {title} {\bibinfo {title} {Scattering and pairing
  in cuprate superconductors},\ }\href@noop {} {\bibfield  {journal} {\bibinfo
  {journal} {Annual Review of Condensed Matter Physics}\ }\textbf {\bibinfo
  {volume} {1}},\ \bibinfo {pages} {51} (\bibinfo {year} {2010})}\BibitemShut
  {NoStop}%
\bibitem [{\citenamefont {Greene}\ \emph {et~al.}(2020)\citenamefont {Greene},
  \citenamefont {Mandal}, \citenamefont {Poniatowski},\ and\ \citenamefont
  {Sarkar}}]{Greene20-Strange-Metal-Electron-doped}%
  \BibitemOpen
  \bibfield  {author} {\bibinfo {author} {\bibfnamefont {R.~L.}\ \bibnamefont
  {Greene}}, \bibinfo {author} {\bibfnamefont {P.~R.}\ \bibnamefont {Mandal}},
  \bibinfo {author} {\bibfnamefont {N.~R.}\ \bibnamefont {Poniatowski}},\ and\
  \bibinfo {author} {\bibfnamefont {T.}~\bibnamefont {Sarkar}},\ }\bibfield
  {title} {\bibinfo {title} {The strange metal state of the electron-doped
  cuprates},\ }\href@noop {} {\bibfield  {journal} {\bibinfo  {journal} {Annual
  Review of Condensed Matter Physics}\ }\textbf {\bibinfo {volume} {11}},\
  \bibinfo {pages} {213} (\bibinfo {year} {2020})}\BibitemShut {NoStop}%
\bibitem [{\citenamefont {Maier}\ \emph {et~al.}(2020)\citenamefont {Maier},
  \citenamefont {Karakuzu},\ and\ \citenamefont
  {Scalapino}}]{Maier20-Overdoped-End-Cuprates-PhaseDiagram}%
  \BibitemOpen
  \bibfield  {author} {\bibinfo {author} {\bibfnamefont {T.~A.}\ \bibnamefont
  {Maier}}, \bibinfo {author} {\bibfnamefont {S.}~\bibnamefont {Karakuzu}},\
  and\ \bibinfo {author} {\bibfnamefont {D.~J.}\ \bibnamefont {Scalapino}},\
  }\bibfield  {title} {\bibinfo {title} {Overdoped end of the cuprate phase
  diagram},\ }\href@noop {} {\bibfield  {journal} {\bibinfo  {journal} {Phys.
  Rev. Research}\ }\textbf {\bibinfo {volume} {2}},\ \bibinfo {pages} {033132}
  (\bibinfo {year} {2020})}\BibitemShut {NoStop}%
\bibitem [{\citenamefont {Yuan}\ \emph {et~al.}(2022)\citenamefont {Yuan},
  \citenamefont {Chen}, \citenamefont {Jiang}, \citenamefont {Feng},
  \citenamefont {Lin}, \citenamefont {Yu}, \citenamefont {He}, \citenamefont
  {Zhang}, \citenamefont {Jiang}, \citenamefont {Zhang} \emph
  {et~al.}}]{Yuan22-Scaling-Strange-Metal-Cuprates}%
  \BibitemOpen
  \bibfield  {author} {\bibinfo {author} {\bibfnamefont {J.}~\bibnamefont
  {Yuan}}, \bibinfo {author} {\bibfnamefont {Q.}~\bibnamefont {Chen}}, \bibinfo
  {author} {\bibfnamefont {K.}~\bibnamefont {Jiang}}, \bibinfo {author}
  {\bibfnamefont {Z.}~\bibnamefont {Feng}}, \bibinfo {author} {\bibfnamefont
  {Z.}~\bibnamefont {Lin}}, \bibinfo {author} {\bibfnamefont {H.}~\bibnamefont
  {Yu}}, \bibinfo {author} {\bibfnamefont {G.}~\bibnamefont {He}}, \bibinfo
  {author} {\bibfnamefont {J.}~\bibnamefont {Zhang}}, \bibinfo {author}
  {\bibfnamefont {X.}~\bibnamefont {Jiang}}, \bibinfo {author} {\bibfnamefont
  {X.}~\bibnamefont {Zhang}}, \emph {et~al.},\ }\bibfield  {title} {\bibinfo
  {title} {Scaling of the strange-metal scattering in unconventional
  superconductors},\ }\href@noop {} {\bibfield  {journal} {\bibinfo  {journal}
  {Nature}\ }\textbf {\bibinfo {volume} {602}},\ \bibinfo {pages} {431}
  (\bibinfo {year} {2022})}\BibitemShut {NoStop}%
\bibitem [{\citenamefont {Honda}\ \emph {et~al.}(2013)\citenamefont {Honda},
  \citenamefont {Maeta}, \citenamefont {Hirose}, \citenamefont {{\=O}nuki},
  \citenamefont {Miyake},\ and\ \citenamefont
  {Settai}}]{Honda13-CeCu2Ge2-Mag-SC}%
  \BibitemOpen
  \bibfield  {author} {\bibinfo {author} {\bibfnamefont {F.}~\bibnamefont
  {Honda}}, \bibinfo {author} {\bibfnamefont {T.}~\bibnamefont {Maeta}},
  \bibinfo {author} {\bibfnamefont {Y.}~\bibnamefont {Hirose}}, \bibinfo
  {author} {\bibfnamefont {Y.}~\bibnamefont {{\=O}nuki}}, \bibinfo {author}
  {\bibfnamefont {A.}~\bibnamefont {Miyake}},\ and\ \bibinfo {author}
  {\bibfnamefont {R.}~\bibnamefont {Settai}},\ }\bibfield  {title} {\bibinfo
  {title} {Magnetism and superconductivity in \textrm{CeCu$_{2}$Ge$_{2}$} under
  high pressures and magnetic fields},\ }\href@noop {} {\bibfield  {journal}
  {\bibinfo  {journal} {J. Korean Phys. Soc.}\ }\textbf {\bibinfo {volume}
  {63}},\ \bibinfo {pages} {345} (\bibinfo {year} {2013})}\BibitemShut
  {NoStop}%
\bibitem [{\citenamefont {Gegenwart}\ \emph {et~al.}(1998)\citenamefont
  {Gegenwart}, \citenamefont {Langhammer}, \citenamefont {Geibel},
  \citenamefont {Helfrich}, \citenamefont {Lang}, \citenamefont {Sparn},
  \citenamefont {Steglich}, \citenamefont {Horn}, \citenamefont {Donnevert},
  \citenamefont {Link},\ and\ \citenamefont
  {Assmus}}]{Gegenwart98-CeCu2Si2-Brakup-HF}%
  \BibitemOpen
  \bibfield  {author} {\bibinfo {author} {\bibfnamefont {P.}~\bibnamefont
  {Gegenwart}}, \bibinfo {author} {\bibfnamefont {C.}~\bibnamefont
  {Langhammer}}, \bibinfo {author} {\bibfnamefont {C.}~\bibnamefont {Geibel}},
  \bibinfo {author} {\bibfnamefont {R.}~\bibnamefont {Helfrich}}, \bibinfo
  {author} {\bibfnamefont {M.}~\bibnamefont {Lang}}, \bibinfo {author}
  {\bibfnamefont {G.}~\bibnamefont {Sparn}}, \bibinfo {author} {\bibfnamefont
  {F.}~\bibnamefont {Steglich}}, \bibinfo {author} {\bibfnamefont
  {R.}~\bibnamefont {Horn}}, \bibinfo {author} {\bibfnamefont {L.}~\bibnamefont
  {Donnevert}}, \bibinfo {author} {\bibfnamefont {A.}~\bibnamefont {Link}},\
  and\ \bibinfo {author} {\bibfnamefont {W.}~\bibnamefont {Assmus}},\
  }\bibfield  {title} {\bibinfo {title} {Breakup of heavy fermions on the brink
  of phase $\mathit{A}$ in \textrm{CeCu$_{2}$Si$_{2}$}},\ }\href@noop {}
  {\bibfield  {journal} {\bibinfo  {journal} {Phys. Rev. Lett.}\ }\textbf
  {\bibinfo {volume} {81}},\ \bibinfo {pages} {1501} (\bibinfo {year}
  {1998})}\BibitemShut {NoStop}%
\bibitem [{\citenamefont {Bellarbi}\ \emph {et~al.}(1984)\citenamefont
  {Bellarbi}, \citenamefont {Benoit}, \citenamefont {Jaccard}, \citenamefont
  {Mignot},\ and\ \citenamefont
  {Braun}}]{Bellarbi84-CeCu2Si2-Valence-Instability}%
  \BibitemOpen
  \bibfield  {author} {\bibinfo {author} {\bibfnamefont {B.}~\bibnamefont
  {Bellarbi}}, \bibinfo {author} {\bibfnamefont {A.}~\bibnamefont {Benoit}},
  \bibinfo {author} {\bibfnamefont {D.}~\bibnamefont {Jaccard}}, \bibinfo
  {author} {\bibfnamefont {J.~M.}\ \bibnamefont {Mignot}},\ and\ \bibinfo
  {author} {\bibfnamefont {H.~F.}\ \bibnamefont {Braun}},\ }\bibfield  {title}
  {\bibinfo {title} {High-pressure valence instability and {$T_{c}$} maximum in
  superconducting \textrm{CeCu$_{2}$Si$_{2}$}},\ }\href@noop {} {\bibfield
  {journal} {\bibinfo  {journal} {Phys. Rev. B}\ }\textbf {\bibinfo {volume}
  {30}},\ \bibinfo {pages} {1182} (\bibinfo {year} {1984})}\BibitemShut
  {NoStop}%
\bibitem [{\citenamefont {Rueff}\ \emph {et~al.}(2011)\citenamefont {Rueff},
  \citenamefont {Raymond}, \citenamefont {Taguchi}, \citenamefont {Sikora},
  \citenamefont {Iti\'e}, \citenamefont {Baudelet}, \citenamefont
  {Braithwaite}, \citenamefont {Knebel},\ and\ \citenamefont
  {Jaccard}}]{Rueff11-CeCu2Si2-VQCP}%
  \BibitemOpen
  \bibfield  {author} {\bibinfo {author} {\bibfnamefont {J.-P.}\ \bibnamefont
  {Rueff}}, \bibinfo {author} {\bibfnamefont {S.}~\bibnamefont {Raymond}},
  \bibinfo {author} {\bibfnamefont {M.}~\bibnamefont {Taguchi}}, \bibinfo
  {author} {\bibfnamefont {M.}~\bibnamefont {Sikora}}, \bibinfo {author}
  {\bibfnamefont {J.-P.}\ \bibnamefont {Iti\'e}}, \bibinfo {author}
  {\bibfnamefont {F.}~\bibnamefont {Baudelet}}, \bibinfo {author}
  {\bibfnamefont {D.}~\bibnamefont {Braithwaite}}, \bibinfo {author}
  {\bibfnamefont {G.}~\bibnamefont {Knebel}},\ and\ \bibinfo {author}
  {\bibfnamefont {D.}~\bibnamefont {Jaccard}},\ }\bibfield  {title} {\bibinfo
  {title} {Pressure-induced valence crossover in superconducting
  \textrm{CeCu$_{2}$Si$_{2}$}},\ }\href@noop {} {\bibfield  {journal} {\bibinfo
   {journal} {Phys. Rev. Lett.}\ }\textbf {\bibinfo {volume} {106}},\ \bibinfo
  {pages} {186405} (\bibinfo {year} {2011})}\BibitemShut {NoStop}%
\bibitem [{\citenamefont {Holmes}\ \emph {et~al.}(2004)\citenamefont {Holmes},
  \citenamefont {Jaccard},\ and\ \citenamefont
  {Miyake}}]{Holmes04-CeCu2Si2-ValenceFluctuation}%
  \BibitemOpen
  \bibfield  {author} {\bibinfo {author} {\bibfnamefont {A.~T.}\ \bibnamefont
  {Holmes}}, \bibinfo {author} {\bibfnamefont {D.}~\bibnamefont {Jaccard}},\
  and\ \bibinfo {author} {\bibfnamefont {K.}~\bibnamefont {Miyake}},\
  }\bibfield  {title} {\bibinfo {title} {Signatures of valence fluctuations in
  \textrm{CeCu$_{2}$Si$_{2}$} under high pressure},\ }\href@noop {} {\bibfield
  {journal} {\bibinfo  {journal} {Phys. Rev. B}\ }\textbf {\bibinfo {volume}
  {69}},\ \bibinfo {pages} {024508} (\bibinfo {year} {2004})}\BibitemShut
  {NoStop}%
\bibitem [{\citenamefont {Lawrence}\ and\ \citenamefont
  {Wilkins}(1973)}]{Lawrence73-ee-Scattering-in-Res}%
  \BibitemOpen
  \bibfield  {author} {\bibinfo {author} {\bibfnamefont {W.~E.}\ \bibnamefont
  {Lawrence}}\ and\ \bibinfo {author} {\bibfnamefont {J.~W.}\ \bibnamefont
  {Wilkins}},\ }\bibfield  {title} {\bibinfo {title} {Electron-electron
  scattering in the transport coefficients of simple metals},\ }\href@noop {}
  {\bibfield  {journal} {\bibinfo  {journal} {Phys. Rev. B}\ }\textbf {\bibinfo
  {volume} {7}},\ \bibinfo {pages} {2317} (\bibinfo {year} {1973})}\BibitemShut
  {NoStop}%
\bibitem [{\citenamefont
  {Lawrence}(1976)}]{Lawrence76-ee-Scattering-NobleMetals}%
  \BibitemOpen
  \bibfield  {author} {\bibinfo {author} {\bibfnamefont {W.~E.}\ \bibnamefont
  {Lawrence}},\ }\bibfield  {title} {\bibinfo {title} {Electron-electron
  scattering in the low-temperature resistivity of the noble metals},\
  }\href@noop {} {\bibfield  {journal} {\bibinfo  {journal} {Phys. Rev. B}\
  }\textbf {\bibinfo {volume} {13}},\ \bibinfo {pages} {5316} (\bibinfo {year}
  {1976})}\BibitemShut {NoStop}%
\bibitem [{\citenamefont
  {MacDonald}(1980)}]{MacDonald80-e-F-enhanced-e-e-Interaction}%
  \BibitemOpen
  \bibfield  {author} {\bibinfo {author} {\bibfnamefont {A.~H.}\ \bibnamefont
  {MacDonald}},\ }\bibfield  {title} {\bibinfo {title} {Electron-phonon
  enhancement of electron-electron scattering in \textrm{Al}},\ }\href@noop {}
  {\bibfield  {journal} {\bibinfo  {journal} {Phys. Rev. Lett.}\ }\textbf
  {\bibinfo {volume} {44}},\ \bibinfo {pages} {489} (\bibinfo {year}
  {1980})}\BibitemShut {NoStop}%
\bibitem [{\citenamefont {MacDonald}\ \emph {et~al.}(1981)\citenamefont
  {MacDonald}, \citenamefont {Taylor},\ and\ \citenamefont
  {Geldart}}]{MacDonald81-Umkalpp-resistivity}%
  \BibitemOpen
  \bibfield  {author} {\bibinfo {author} {\bibfnamefont {A.~H.}\ \bibnamefont
  {MacDonald}}, \bibinfo {author} {\bibfnamefont {R.}~\bibnamefont {Taylor}},\
  and\ \bibinfo {author} {\bibfnamefont {D.~J.~W.}\ \bibnamefont {Geldart}},\
  }\bibfield  {title} {\bibinfo {title} {Umklapp electron-electron scattering
  and the low-temperature electrical resistivity of the alkali metals},\
  }\href@noop {} {\bibfield  {journal} {\bibinfo  {journal} {Phys. Rev. B}\
  }\textbf {\bibinfo {volume} {23}},\ \bibinfo {pages} {2718} (\bibinfo {year}
  {1981})}\BibitemShut {NoStop}%
\bibitem [{\citenamefont {Patton}\ and\ \citenamefont
  {Zaringhalam}(1975)}]{Patton75-FermiLiquid-Superfluid}%
  \BibitemOpen
  \bibfield  {author} {\bibinfo {author} {\bibfnamefont {B.}~\bibnamefont
  {Patton}}\ and\ \bibinfo {author} {\bibfnamefont {A.}~\bibnamefont
  {Zaringhalam}},\ }\bibfield  {title} {\bibinfo {title} {The superfluidd
  transition temperature of a {F}ermi liquid: $^{3}${H}e and
  $^{3}${H}e-$^{4}${H}e mixtures},\ }\href@noop {} {\bibfield  {journal}
  {\bibinfo  {journal} {Phys. Lett. A}\ }\textbf {\bibinfo {volume} {55}},\
  \bibinfo {pages} {95} (\bibinfo {year} {1975})}\BibitemShut {NoStop}%
\bibitem [{\citenamefont {Pethick}\ \emph {et~al.}(1986)\citenamefont
  {Pethick}, \citenamefont {Pines}, \citenamefont {Quader}, \citenamefont
  {Bedell},\ and\ \citenamefont {Brown}}]{Pethick86-FL-Theory-UPt3}%
  \BibitemOpen
  \bibfield  {author} {\bibinfo {author} {\bibfnamefont {C.~J.}\ \bibnamefont
  {Pethick}}, \bibinfo {author} {\bibfnamefont {D.}~\bibnamefont {Pines}},
  \bibinfo {author} {\bibfnamefont {K.~F.}\ \bibnamefont {Quader}}, \bibinfo
  {author} {\bibfnamefont {K.~S.}\ \bibnamefont {Bedell}},\ and\ \bibinfo
  {author} {\bibfnamefont {G.~E.}\ \bibnamefont {Brown}},\ }\bibfield  {title}
  {\bibinfo {title} {One-component {F}ermi-liquid theory and the properties of
  \textrm{UPt$_{3}$}},\ }\href@noop {} {\bibfield  {journal} {\bibinfo
  {journal} {Phys. Rev. Lett.}\ }\textbf {\bibinfo {volume} {57}},\ \bibinfo
  {pages} {1955} (\bibinfo {year} {1986})}\BibitemShut {NoStop}%
\bibitem [{\citenamefont {Miranda}\ and\ \citenamefont
  {Dobrosavljevi{\'c}}(2005)}]{Miranda05-disorder-driven-NFL-Review}%
  \BibitemOpen
  \bibfield  {author} {\bibinfo {author} {\bibfnamefont {E.}~\bibnamefont
  {Miranda}}\ and\ \bibinfo {author} {\bibfnamefont {V.}~\bibnamefont
  {Dobrosavljevi{\'c}}},\ }\bibfield  {title} {\bibinfo {title}
  {Disorder-driven non-{F}ermi liquid behaviour of correlated electrons},\
  }\href@noop {} {\bibfield  {journal} {\bibinfo  {journal} {Rep. on Prog. in
  Phys.}\ }\textbf {\bibinfo {volume} {68}},\ \bibinfo {pages} {2337} (\bibinfo
  {year} {2005})}\BibitemShut {NoStop}%
\bibitem [{\citenamefont {Sheikin}\ \emph {et~al.}(2000)\citenamefont
  {Sheikin}, \citenamefont {Braithwaite}, \citenamefont {Brison}, \citenamefont
  {Assmus},\ and\ \citenamefont {Flouquet}}]{Sheikin00-CeCu2Si2-HF-R-P}%
  \BibitemOpen
  \bibfield  {author} {\bibinfo {author} {\bibfnamefont {I.}~\bibnamefont
  {Sheikin}}, \bibinfo {author} {\bibfnamefont {D.}~\bibnamefont
  {Braithwaite}}, \bibinfo {author} {\bibfnamefont {J.-P.}\ \bibnamefont
  {Brison}}, \bibinfo {author} {\bibfnamefont {W.}~\bibnamefont {Assmus}},\
  and\ \bibinfo {author} {\bibfnamefont {J.}~\bibnamefont {Flouquet}},\
  }\bibfield  {title} {\bibinfo {title} {Transport measurements of the heavy
  fermion superconductor \textrm{CeCu$_{2}$Si$_{2}$} under pressure},\
  }\href@noop {} {\bibfield  {journal} {\bibinfo  {journal} {J. Low Temp.
  Phys.}\ }\textbf {\bibinfo {volume} {118}},\ \bibinfo {pages} {113} (\bibinfo
  {year} {2000})}\BibitemShut {NoStop}%
\bibitem [{\citenamefont {Rosch}(1999)}]{Rosch99-disorder-SpinFluctuation-QCP}%
  \BibitemOpen
  \bibfield  {author} {\bibinfo {author} {\bibfnamefont {A.}~\bibnamefont
  {Rosch}},\ }\bibfield  {title} {\bibinfo {title} {Interplay of disorder and
  spin fluctuations in the resistivity near a quantum critical point},\
  }\href@noop {} {\bibfield  {journal} {\bibinfo  {journal} {Phys. Rev. Lett.}\
  }\textbf {\bibinfo {volume} {82}},\ \bibinfo {pages} {4280} (\bibinfo {year}
  {1999})}\BibitemShut {NoStop}%
\bibitem [{\citenamefont {Kambe}\ and\ \citenamefont
  {Flouquet}(1997)}]{Kambe97-CeCu2Si2-Purity}%
  \BibitemOpen
  \bibfield  {author} {\bibinfo {author} {\bibfnamefont {S.}~\bibnamefont
  {Kambe}}\ and\ \bibinfo {author} {\bibfnamefont {J.}~\bibnamefont
  {Flouquet}},\ }\bibfield  {title} {\bibinfo {title} {Kondo impurity
  scattering and the magnetic quantum critical transition},\ }\href@noop {}
  {\bibfield  {journal} {\bibinfo  {journal} {Solid State Commun.}\ }\textbf
  {\bibinfo {volume} {103}},\ \bibinfo {pages} {551} (\bibinfo {year}
  {1997})}\BibitemShut {NoStop}%
\bibitem [{\citenamefont {Sidorov}\ \emph {et~al.}(2002)\citenamefont
  {Sidorov}, \citenamefont {Nicklas}, \citenamefont {Pagliuso}, \citenamefont
  {Sarrao}, \citenamefont {Bang}, \citenamefont {Balatsky},\ and\ \citenamefont
  {Thompson}}]{Sidorov02-CeCoIn5-Q-Criticality}%
  \BibitemOpen
  \bibfield  {author} {\bibinfo {author} {\bibfnamefont {V.}~\bibnamefont
  {Sidorov}}, \bibinfo {author} {\bibfnamefont {M.}~\bibnamefont {Nicklas}},
  \bibinfo {author} {\bibfnamefont {P.}~\bibnamefont {Pagliuso}}, \bibinfo
  {author} {\bibfnamefont {J.}~\bibnamefont {Sarrao}}, \bibinfo {author}
  {\bibfnamefont {Y.}~\bibnamefont {Bang}}, \bibinfo {author} {\bibfnamefont
  {A.~V.}\ \bibnamefont {Balatsky}},\ and\ \bibinfo {author} {\bibfnamefont
  {J.~D.}\ \bibnamefont {Thompson}},\ }\bibfield  {title} {\bibinfo {title}
  {Superconductivity and quantum criticality in \ce{CeCoIn5}},\ }\href@noop {}
  {\bibfield  {journal} {\bibinfo  {journal} {Phys. Rev. Lett.}\ }\textbf
  {\bibinfo {volume} {89}},\ \bibinfo {pages} {157004} (\bibinfo {year}
  {2002})}\BibitemShut {NoStop}%
\bibitem [{\citenamefont {Kotegawa}\ \emph {et~al.}(2006)\citenamefont
  {Kotegawa}, \citenamefont {Takeda}, \citenamefont {Miyoshi}, \citenamefont
  {Fukushima}, \citenamefont {Hidaka}, \citenamefont {C.~Kobayashi},
  \citenamefont {Akazawa}, \citenamefont {Ohishi}, \citenamefont {Nakashima},
  \citenamefont {Thamizhavel} \emph {et~al.}}]{Kotegawa06-CeNiG3-SC-AFM}%
  \BibitemOpen
  \bibfield  {author} {\bibinfo {author} {\bibfnamefont {H.}~\bibnamefont
  {Kotegawa}}, \bibinfo {author} {\bibfnamefont {K.}~\bibnamefont {Takeda}},
  \bibinfo {author} {\bibfnamefont {T.}~\bibnamefont {Miyoshi}}, \bibinfo
  {author} {\bibfnamefont {S.}~\bibnamefont {Fukushima}}, \bibinfo {author}
  {\bibfnamefont {H.}~\bibnamefont {Hidaka}}, \bibinfo {author} {\bibfnamefont
  {T.}~\bibnamefont {C.~Kobayashi}}, \bibinfo {author} {\bibfnamefont
  {T.}~\bibnamefont {Akazawa}}, \bibinfo {author} {\bibfnamefont
  {Y.}~\bibnamefont {Ohishi}}, \bibinfo {author} {\bibfnamefont
  {M.}~\bibnamefont {Nakashima}}, \bibinfo {author} {\bibfnamefont
  {A.}~\bibnamefont {Thamizhavel}}, \emph {et~al.},\ }\bibfield  {title}
  {\bibinfo {title} {Pressure-induced superconductivity emerging from
  antiferromagnetic phase in \textrm{CeNiGe$_3$}},\ }\href@noop {} {\bibfield
  {journal} {\bibinfo  {journal} {J. Phys. Soc. Jpn.}\ }\textbf {\bibinfo
  {volume} {75}},\ \bibinfo {pages} {044713} (\bibinfo {year}
  {2006})}\BibitemShut {NoStop}%
\bibitem [{\citenamefont {Araki}\ \emph {et~al.}(2002)\citenamefont {Araki},
  \citenamefont {Nakashima}, \citenamefont {Settai}, \citenamefont
  {Kobayashi},\ and\ \citenamefont {Onuki}}]{Araki02-CeRh2Si2-SC-pressure}%
  \BibitemOpen
  \bibfield  {author} {\bibinfo {author} {\bibfnamefont {S.}~\bibnamefont
  {Araki}}, \bibinfo {author} {\bibfnamefont {M.}~\bibnamefont {Nakashima}},
  \bibinfo {author} {\bibfnamefont {R.}~\bibnamefont {Settai}}, \bibinfo
  {author} {\bibfnamefont {T.~C.}\ \bibnamefont {Kobayashi}},\ and\ \bibinfo
  {author} {\bibfnamefont {Y.}~\bibnamefont {Onuki}},\ }\bibfield  {title}
  {\bibinfo {title} {Pressure-induced superconductivity in an antiferromagnet
  \textrm{CeRh$_{2}$Si$_{2}$}},\ }\href@noop {} {\bibfield  {journal} {\bibinfo
   {journal} {J. Phys.: Condens. Matter}\ }\textbf {\bibinfo {volume} {14}},\
  \bibinfo {pages} {L377} (\bibinfo {year} {2002})}\BibitemShut {NoStop}%
\bibitem [{\citenamefont {Nakashima}\ \emph {et~al.}(2004)\citenamefont
  {Nakashima}, \citenamefont {Tabata}, \citenamefont {Thamizhavel},
  \citenamefont {Kobayashi}, \citenamefont {Hedo}, \citenamefont {Uwatoko},
  \citenamefont {Shimizu}, \citenamefont {Settai},\ and\ \citenamefont
  {Onuki}}]{Nakashima04-CeNiGe3}%
  \BibitemOpen
  \bibfield  {author} {\bibinfo {author} {\bibfnamefont {M.}~\bibnamefont
  {Nakashima}}, \bibinfo {author} {\bibfnamefont {K.}~\bibnamefont {Tabata}},
  \bibinfo {author} {\bibfnamefont {A.}~\bibnamefont {Thamizhavel}}, \bibinfo
  {author} {\bibfnamefont {T.~C.}\ \bibnamefont {Kobayashi}}, \bibinfo {author}
  {\bibfnamefont {M.}~\bibnamefont {Hedo}}, \bibinfo {author} {\bibfnamefont
  {Y.}~\bibnamefont {Uwatoko}}, \bibinfo {author} {\bibfnamefont
  {K.}~\bibnamefont {Shimizu}}, \bibinfo {author} {\bibfnamefont
  {R.}~\bibnamefont {Settai}},\ and\ \bibinfo {author} {\bibfnamefont
  {Y.}~\bibnamefont {Onuki}},\ }\bibfield  {title} {\bibinfo {title}
  {High-pressure effect on the electronic state in \ce{CeNiGe3} :
  pressure-induced superconductivity},\ }\href@noop {} {\bibfield  {journal}
  {\bibinfo  {journal} {J. Phys.: Condens. Matter}\ }\textbf {\bibinfo {volume}
  {16}},\ \bibinfo {pages} {L255} (\bibinfo {year} {2004})}\BibitemShut
  {NoStop}%
\bibitem [{\citenamefont {Nakashima}\ \emph {et~al.}(2006)\citenamefont
  {Nakashima}, \citenamefont {Kohara}, \citenamefont {Thamizhavel},
  \citenamefont {Matsuda}, \citenamefont {Haga}, \citenamefont {Hedo},
  \citenamefont {Uwatoko}, \citenamefont {Settai},\ and\ \citenamefont
  {{\=O}nuki}}]{Nakashima06-Ce2Ni3Ge5-SC-Pressure}%
  \BibitemOpen
  \bibfield  {author} {\bibinfo {author} {\bibfnamefont {M.}~\bibnamefont
  {Nakashima}}, \bibinfo {author} {\bibfnamefont {H.}~\bibnamefont {Kohara}},
  \bibinfo {author} {\bibfnamefont {A.}~\bibnamefont {Thamizhavel}}, \bibinfo
  {author} {\bibfnamefont {T.~D.}\ \bibnamefont {Matsuda}}, \bibinfo {author}
  {\bibfnamefont {Y.}~\bibnamefont {Haga}}, \bibinfo {author} {\bibfnamefont
  {M.}~\bibnamefont {Hedo}}, \bibinfo {author} {\bibfnamefont {Y.}~\bibnamefont
  {Uwatoko}}, \bibinfo {author} {\bibfnamefont {R.}~\bibnamefont {Settai}},\
  and\ \bibinfo {author} {\bibfnamefont {Y.}~\bibnamefont {{\=O}nuki}},\
  }\bibfield  {title} {\bibinfo {title} {Pressure-induced superconductivity of
  \textrm{Ce$_{2}$Ni$_{3}$Ge$_{5}$}},\ }\href@noop {} {\bibfield  {journal}
  {\bibinfo  {journal} {Physica B: Cond Matter}\ }\textbf {\bibinfo {volume}
  {378}},\ \bibinfo {pages} {402} (\bibinfo {year} {2006})}\BibitemShut
  {NoStop}%
\bibitem [{\citenamefont {Onuki}\ \emph {et~al.}(2008)\citenamefont {Onuki},
  \citenamefont {Miyauchi}, \citenamefont {Tsujino}, \citenamefont {Ida},
  \citenamefont {Settai}, \citenamefont {Takeuchi}, \citenamefont {Tateiwa},
  \citenamefont {D.~Matsuda}, \citenamefont {Haga},\ and\ \citenamefont
  {Harima}}]{Onuki08-CePt3Si-CeIr3Si}%
  \BibitemOpen
  \bibfield  {author} {\bibinfo {author} {\bibfnamefont {Y.}~\bibnamefont
  {Onuki}}, \bibinfo {author} {\bibfnamefont {Y.}~\bibnamefont {Miyauchi}},
  \bibinfo {author} {\bibfnamefont {M.}~\bibnamefont {Tsujino}}, \bibinfo
  {author} {\bibfnamefont {Y.}~\bibnamefont {Ida}}, \bibinfo {author}
  {\bibfnamefont {R.}~\bibnamefont {Settai}}, \bibinfo {author} {\bibfnamefont
  {T.}~\bibnamefont {Takeuchi}}, \bibinfo {author} {\bibfnamefont
  {N.}~\bibnamefont {Tateiwa}}, \bibinfo {author} {\bibfnamefont
  {T.}~\bibnamefont {D.~Matsuda}}, \bibinfo {author} {\bibfnamefont
  {Y.}~\bibnamefont {Haga}},\ and\ \bibinfo {author} {\bibfnamefont
  {H.}~\bibnamefont {Harima}},\ }\bibfield  {title} {\bibinfo {title}
  {Superconducting properties of \textrm{CePt$_3$Si} and \textrm{CeIrSi$_3$}
  without inversion symmetry in the crystal structure},\ }\href@noop {}
  {\bibfield  {journal} {\bibinfo  {journal} {J. Phys. Soc. Jpn.}\ }\textbf
  {\bibinfo {volume} {77}},\ \bibinfo {pages} {37} (\bibinfo {year}
  {2008})}\BibitemShut {NoStop}%
\bibitem [{\citenamefont {Hassinger}\ \emph {et~al.}(2008)\citenamefont
  {Hassinger}, \citenamefont {Knebel}, \citenamefont {Izawa}, \citenamefont
  {Lejay}, \citenamefont {Salce},\ and\ \citenamefont
  {Flouquet}}]{Hassinger08-URu2Si2-HiddenOrder-Nesting}%
  \BibitemOpen
  \bibfield  {author} {\bibinfo {author} {\bibfnamefont {E.}~\bibnamefont
  {Hassinger}}, \bibinfo {author} {\bibfnamefont {G.}~\bibnamefont {Knebel}},
  \bibinfo {author} {\bibfnamefont {K.}~\bibnamefont {Izawa}}, \bibinfo
  {author} {\bibfnamefont {P.}~\bibnamefont {Lejay}}, \bibinfo {author}
  {\bibfnamefont {B.}~\bibnamefont {Salce}},\ and\ \bibinfo {author}
  {\bibfnamefont {J.}~\bibnamefont {Flouquet}},\ }\bibfield  {title} {\bibinfo
  {title} {Temperature-pressure phase diagram of \textrm{URu$_{2}$Si$_{2}$}
  from resistivity measurements and ac calorimetry: Hidden order and
  fermi-surface nesting},\ }\href@noop {} {\bibfield  {journal} {\bibinfo
  {journal} {Phys. Rev. B}\ }\textbf {\bibinfo {volume} {77}},\ \bibinfo
  {pages} {115117} (\bibinfo {year} {2008})}\BibitemShut {NoStop}%
\bibitem [{\citenamefont {de~Visser}\ \emph {et~al.}(1984)\citenamefont
  {de~Visser}, \citenamefont {Franse},\ and\ \citenamefont
  {Menovsky}}]{deVisser84-UPt3-FL}%
  \BibitemOpen
  \bibfield  {author} {\bibinfo {author} {\bibfnamefont {A.}~\bibnamefont
  {de~Visser}}, \bibinfo {author} {\bibfnamefont {J.}~\bibnamefont {Franse}},\
  and\ \bibinfo {author} {\bibfnamefont {A.}~\bibnamefont {Menovsky}},\
  }\bibfield  {title} {\bibinfo {title} {Resistivity of single-crystalline
  \ce{UPt3} and its pressure dependence; interpretation by a spin-fluctuation
  model},\ }\href@noop {} {\bibfield  {journal} {\bibinfo  {journal} {J. Magn.
  Magn. Mater.}\ }\textbf {\bibinfo {volume} {43}},\ \bibinfo {pages} {43}
  (\bibinfo {year} {1984})}\BibitemShut {NoStop}%
\bibitem [{\citenamefont {Ott}\ \emph {et~al.}(1983)\citenamefont {Ott},
  \citenamefont {Rudigier}, \citenamefont {Fisk},\ and\ \citenamefont
  {Smith}}]{Ott83-UBe13-Unconv-SC}%
  \BibitemOpen
  \bibfield  {author} {\bibinfo {author} {\bibfnamefont {H.~R.}\ \bibnamefont
  {Ott}}, \bibinfo {author} {\bibfnamefont {H.}~\bibnamefont {Rudigier}},
  \bibinfo {author} {\bibfnamefont {Z.}~\bibnamefont {Fisk}},\ and\ \bibinfo
  {author} {\bibfnamefont {J.~L.}\ \bibnamefont {Smith}},\ }\bibfield  {title}
  {\bibinfo {title} {\ce{UBe13}: An unconventional actinide superconductor},\
  }\href@noop {} {\bibfield  {journal} {\bibinfo  {journal} {Phys. Rev. Lett.}\
  }\textbf {\bibinfo {volume} {50}},\ \bibinfo {pages} {1595} (\bibinfo {year}
  {1983})}\BibitemShut {NoStop}%
\bibitem [{\citenamefont {Nunez-Regueiro}\ \emph {et~al.}(2012)\citenamefont
  {Nunez-Regueiro}, \citenamefont {Garbarino},\ and\ \citenamefont
  {Nunez-Regueiro}}]{Nunes12-FermiLiquid-SUC}%
  \BibitemOpen
  \bibfield  {author} {\bibinfo {author} {\bibfnamefont {M.}~\bibnamefont
  {Nunez-Regueiro}}, \bibinfo {author} {\bibfnamefont {G.}~\bibnamefont
  {Garbarino}},\ and\ \bibinfo {author} {\bibfnamefont {M.~D.}\ \bibnamefont
  {Nunez-Regueiro}},\ }\bibfield  {title} {\bibinfo {title} {The relationship
  between the normal state {F}ermi liquid scattering rate and the
  superconducting state},\ }\href@noop {} {\bibfield  {journal} {\bibinfo
  {journal} {J Phys: Conf. Series}\ }\textbf {\bibinfo {volume} {400}},\
  \bibinfo {pages} {022085} (\bibinfo {year} {2012})}\BibitemShut {NoStop}%
\bibitem [{\citenamefont {Castro}\ \emph {et~al.}(2018)\citenamefont {Castro},
  \citenamefont {Ferreira}, \citenamefont {Neto},\ and\ \citenamefont
  {ElMassalami}}]{18-Castro-Tc-A-Correlation}%
  \BibitemOpen
  \bibfield  {author} {\bibinfo {author} {\bibfnamefont {P.~B.}\ \bibnamefont
  {Castro}}, \bibinfo {author} {\bibfnamefont {J.~L.}\ \bibnamefont
  {Ferreira}}, \bibinfo {author} {\bibfnamefont {M.~B.~S.}\ \bibnamefont
  {Neto}},\ and\ \bibinfo {author} {\bibfnamefont {M.}~\bibnamefont
  {ElMassalami}},\ }\bibfield  {title} {\bibinfo {title} {Correlation of
  {$T_{c}$} and coefficient of \textrm{T$^{2}$} resistivity term of {Fe}-based
  pnictide \& chalcogenide superconductors},\ }\href@noop {} {\bibfield
  {journal} {\bibinfo  {journal} {J. Phys.: Conf. Ser.}\ }\textbf {\bibinfo
  {volume} {969}},\ \bibinfo {pages} {012050} (\bibinfo {year}
  {2018})}\BibitemShut {NoStop}%
\bibitem [{\citenamefont {Soares}\ \emph {et~al.}(2018)\citenamefont {Soares},
  \citenamefont {ElMassalami}, \citenamefont {Yanagisawa}, \citenamefont
  {Tanaka}, \citenamefont {Takeya},\ and\ \citenamefont
  {Takano}}]{18-Soares-KxFe2-ySe2-Quantum-Conductance-Phase-Diagram}%
  \BibitemOpen
  \bibfield  {author} {\bibinfo {author} {\bibfnamefont {C.}~\bibnamefont
  {Soares}}, \bibinfo {author} {\bibfnamefont {M.}~\bibnamefont {ElMassalami}},
  \bibinfo {author} {\bibfnamefont {Y.}~\bibnamefont {Yanagisawa}}, \bibinfo
  {author} {\bibfnamefont {M.}~\bibnamefont {Tanaka}}, \bibinfo {author}
  {\bibfnamefont {H.}~\bibnamefont {Takeya}},\ and\ \bibinfo {author}
  {\bibfnamefont {Y.}~\bibnamefont {Takano}},\ }\bibfield  {title} {\bibinfo
  {title} {Quantum conductance-temperature phase diagram of granular
  superconductor \textrm{K$_{x}$Fe$_{2-y}$Se$_{2}$}},\ }\href@noop {}
  {\bibfield  {journal} {\bibinfo  {journal} {Sci. Rep.}\ }\textbf {\bibinfo
  {volume} {8}},\ \bibinfo {pages} {7041} (\bibinfo {year} {2018})}\BibitemShut
  {NoStop}%
\bibitem [{\citenamefont
  {Gurvitch}(1986)}]{Gurvitch86-Disorder-induced-transition-AT2}%
  \BibitemOpen
  \bibfield  {author} {\bibinfo {author} {\bibfnamefont {M.}~\bibnamefont
  {Gurvitch}},\ }\bibfield  {title} {\bibinfo {title} {Universal
  disorder-induced transition in the resistivity behavior of strongly coupled
  metals},\ }\href@noop {} {\bibfield  {journal} {\bibinfo  {journal} {Phys.
  Rev. Lett.}\ }\textbf {\bibinfo {volume} {56}},\ \bibinfo {pages} {647}
  (\bibinfo {year} {1986})}\BibitemShut {NoStop}%
\bibitem [{\citenamefont {Ronning}\ \emph {et~al.}(2006)\citenamefont
  {Ronning}, \citenamefont {Capan}, \citenamefont {Bauer}, \citenamefont
  {Thompson}, \citenamefont {Sarrao},\ and\ \citenamefont
  {Movshovich}}]{Ronning06-CeCoIn5-Pressure-QCP}%
  \BibitemOpen
  \bibfield  {author} {\bibinfo {author} {\bibfnamefont {F.}~\bibnamefont
  {Ronning}}, \bibinfo {author} {\bibfnamefont {C.}~\bibnamefont {Capan}},
  \bibinfo {author} {\bibfnamefont {E.~D.}\ \bibnamefont {Bauer}}, \bibinfo
  {author} {\bibfnamefont {J.~D.}\ \bibnamefont {Thompson}}, \bibinfo {author}
  {\bibfnamefont {J.~L.}\ \bibnamefont {Sarrao}},\ and\ \bibinfo {author}
  {\bibfnamefont {R.}~\bibnamefont {Movshovich}},\ }\bibfield  {title}
  {\bibinfo {title} {Pressure study of quantum criticality in
  \textrm{CeCoIn$_{5}$}},\ }\href@noop {} {\bibfield  {journal} {\bibinfo
  {journal} {Phys. Rev. B}\ }\textbf {\bibinfo {volume} {73}},\ \bibinfo
  {pages} {064519} (\bibinfo {year} {2006})}\BibitemShut {NoStop}%
\bibitem [{\citenamefont {Paglione}\ \emph {et~al.}(2003)\citenamefont
  {Paglione}, \citenamefont {Tanatar}, \citenamefont {Hawthorn}, \citenamefont
  {Boaknin}, \citenamefont {Hill}, \citenamefont {Ronning}, \citenamefont
  {Sutherland}, \citenamefont {Taillefer}, \citenamefont {Petrovic},\ and\
  \citenamefont {Canfield}}]{Paglione03-CeCoIn5-H-QCP}%
  \BibitemOpen
  \bibfield  {author} {\bibinfo {author} {\bibfnamefont {J.}~\bibnamefont
  {Paglione}}, \bibinfo {author} {\bibfnamefont {M.~A.}\ \bibnamefont
  {Tanatar}}, \bibinfo {author} {\bibfnamefont {D.~G.}\ \bibnamefont
  {Hawthorn}}, \bibinfo {author} {\bibfnamefont {E.}~\bibnamefont {Boaknin}},
  \bibinfo {author} {\bibfnamefont {R.~W.}\ \bibnamefont {Hill}}, \bibinfo
  {author} {\bibfnamefont {F.}~\bibnamefont {Ronning}}, \bibinfo {author}
  {\bibfnamefont {M.}~\bibnamefont {Sutherland}}, \bibinfo {author}
  {\bibfnamefont {L.}~\bibnamefont {Taillefer}}, \bibinfo {author}
  {\bibfnamefont {C.}~\bibnamefont {Petrovic}},\ and\ \bibinfo {author}
  {\bibfnamefont {P.~C.}\ \bibnamefont {Canfield}},\ }\bibfield  {title}
  {\bibinfo {title} {Field-induced quantum critical point in
  \textrm{CeCOIn$_{5}$}},\ }\href@noop {} {\bibfield  {journal} {\bibinfo
  {journal} {Phys. Rev. Lett.}\ }\textbf {\bibinfo {volume} {91}},\ \bibinfo
  {pages} {246405} (\bibinfo {year} {2003})}\BibitemShut {NoStop}%
\bibitem [{\citenamefont {Bianchi}\ \emph {et~al.}(2003)\citenamefont
  {Bianchi}, \citenamefont {Movshovich}, \citenamefont {Vekhter}, \citenamefont
  {Pagliuso},\ and\ \citenamefont {Sarrao}}]{Bianchi03-CeCoIn5-H-QCP}%
  \BibitemOpen
  \bibfield  {author} {\bibinfo {author} {\bibfnamefont {A.}~\bibnamefont
  {Bianchi}}, \bibinfo {author} {\bibfnamefont {R.}~\bibnamefont {Movshovich}},
  \bibinfo {author} {\bibfnamefont {I.}~\bibnamefont {Vekhter}}, \bibinfo
  {author} {\bibfnamefont {P.~G.}\ \bibnamefont {Pagliuso}},\ and\ \bibinfo
  {author} {\bibfnamefont {J.~L.}\ \bibnamefont {Sarrao}},\ }\bibfield  {title}
  {\bibinfo {title} {Avoided antiferromagnetic order and quantum critical point
  in
  $\mathrm{C}\mathrm{e}\mathrm{C}\mathrm{o}\mathrm{I}{\mathrm{n}}_{\mathrm{5}}$},\
  }\href@noop {} {\bibfield  {journal} {\bibinfo  {journal} {Phys. Rev. Lett.}\
  }\textbf {\bibinfo {volume} {91}},\ \bibinfo {pages} {257001} (\bibinfo
  {year} {2003})}\BibitemShut {NoStop}%
\bibitem [{\citenamefont {Sachdev}(1999)}]{Sachdev99-Quantum-Phase-Transition}%
  \BibitemOpen
  \bibfield  {author} {\bibinfo {author} {\bibfnamefont {S.}~\bibnamefont
  {Sachdev}},\ }\bibfield  {title} {\bibinfo {title} {Quantum phase
  transitions},\ }\href@noop {} {\bibfield  {journal} {\bibinfo  {journal}
  {Physics world}\ }\textbf {\bibinfo {volume} {12}},\ \bibinfo {pages} {33}
  (\bibinfo {year} {1999})}\BibitemShut {NoStop}%
\bibitem [{\citenamefont
  {Millis}(1993)}]{Millis93-nonZero-Temperature-on-QCP-Itinerat-Fermi}%
  \BibitemOpen
  \bibfield  {author} {\bibinfo {author} {\bibfnamefont {A.~J.}\ \bibnamefont
  {Millis}},\ }\bibfield  {title} {\bibinfo {title} {Effect of a nonzero
  temperature on quantum critical points in itinerant fermion systems},\
  }\href@noop {} {\bibfield  {journal} {\bibinfo  {journal} {Phys. Rev. B}\
  }\textbf {\bibinfo {volume} {48}},\ \bibinfo {pages} {7183} (\bibinfo {year}
  {1993})}\BibitemShut {NoStop}%
\bibitem [{\citenamefont {Allen}\ and\ \citenamefont
  {Dynes}(1975)}]{Allen-Dynes75-Tc-S-C}%
  \BibitemOpen
  \bibfield  {author} {\bibinfo {author} {\bibfnamefont {P.~B.}\ \bibnamefont
  {Allen}}\ and\ \bibinfo {author} {\bibfnamefont {R.~C.}\ \bibnamefont
  {Dynes}},\ }\bibfield  {title} {\bibinfo {title} {Tc of strong-coupled
  superconductors reanalyzed},\ }\href@noop {} {\bibfield  {journal} {\bibinfo
  {journal} {Phys. Rev. B}\ }\textbf {\bibinfo {volume} {12}},\ \bibinfo
  {pages} {905} (\bibinfo {year} {1975})}\BibitemShut {NoStop}%
\bibitem [{\citenamefont
  {Carbotte}(1990)}]{Carbotte90-SCs-Boson-Exchanged-Reivew}%
  \BibitemOpen
  \bibfield  {author} {\bibinfo {author} {\bibfnamefont {J.~P.}\ \bibnamefont
  {Carbotte}},\ }\bibfield  {title} {\bibinfo {title} {Properties of
  boson-exchange superconductors},\ }\href@noop {} {\bibfield  {journal}
  {\bibinfo  {journal} {Rev. Mod. Phys.}\ }\textbf {\bibinfo {volume} {62}},\
  \bibinfo {pages} {1027} (\bibinfo {year} {1990})}\BibitemShut {NoStop}%
\bibitem [{\citenamefont {McMillan}(1968)}]{Mcmillan68-Tc}%
  \BibitemOpen
  \bibfield  {author} {\bibinfo {author} {\bibfnamefont {W.~L.}\ \bibnamefont
  {McMillan}},\ }\bibfield  {title} {\bibinfo {title} {Tc of strong-coiupled
  superconductors},\ }\href@noop {} {\bibfield  {journal} {\bibinfo  {journal}
  {Phys. Rev.}\ }\textbf {\bibinfo {volume} {167}},\ \bibinfo {pages} {331}
  (\bibinfo {year} {1968})}\BibitemShut {NoStop}%
\bibitem [{\citenamefont {Tsai}\ \emph {et~al.}(2005)\citenamefont {Tsai},
  \citenamefont {Castro~Neto}, \citenamefont {Shankar},\ and\ \citenamefont
  {Campbell}}]{Tsai05-Renormalization-Strong-coupled-SUCs}%
  \BibitemOpen
  \bibfield  {author} {\bibinfo {author} {\bibfnamefont {S.-W.}\ \bibnamefont
  {Tsai}}, \bibinfo {author} {\bibfnamefont {A.~H.}\ \bibnamefont
  {Castro~Neto}}, \bibinfo {author} {\bibfnamefont {R.}~\bibnamefont
  {Shankar}},\ and\ \bibinfo {author} {\bibfnamefont {D.~K.}\ \bibnamefont
  {Campbell}},\ }\bibfield  {title} {\bibinfo {title} {Renormalization-group
  approach to strong-coupled superconductors},\ }\href@noop {} {\bibfield
  {journal} {\bibinfo  {journal} {Phys. Rev. B}\ }\textbf {\bibinfo {volume}
  {72}},\ \bibinfo {pages} {054531} (\bibinfo {year} {2005})}\BibitemShut
  {NoStop}%
\bibitem [{\citenamefont {ElMassalami}\ and\ \citenamefont
  {Neto}(2021)}]{21-FL-SC-Defectal-Reconciled}%
  \BibitemOpen
  \bibfield  {author} {\bibinfo {author} {\bibfnamefont {M.}~\bibnamefont
  {ElMassalami}}\ and\ \bibinfo {author} {\bibfnamefont {M.~B.~S.}\
  \bibnamefont {Neto}},\ }\bibfield  {title} {\bibinfo {title}
  {Superconductivity, fermi-liquid transport, and universal kinematic scaling
  relation for metallic thin films with stabilized defect complexes},\
  }\href@noop {} {\bibfield  {journal} {\bibinfo  {journal} {Phys. Rev. B}\
  }\textbf {\bibinfo {volume} {104}},\ \bibinfo {pages} {014520} (\bibinfo
  {year} {2021})}\BibitemShut {NoStop}%
\bibitem [{\citenamefont {Bauer}\ \emph {et~al.}(2005)\citenamefont {Bauer},
  \citenamefont {Capan}, \citenamefont {Ronning}, \citenamefont {Movshovich},
  \citenamefont {Thompson},\ and\ \citenamefont
  {Sarrao}}]{Bauer05-CeCoIn5-xSn-x-HCP}%
  \BibitemOpen
  \bibfield  {author} {\bibinfo {author} {\bibfnamefont {E.~D.}\ \bibnamefont
  {Bauer}}, \bibinfo {author} {\bibfnamefont {C.}~\bibnamefont {Capan}},
  \bibinfo {author} {\bibfnamefont {F.}~\bibnamefont {Ronning}}, \bibinfo
  {author} {\bibfnamefont {R.}~\bibnamefont {Movshovich}}, \bibinfo {author}
  {\bibfnamefont {J.~D.}\ \bibnamefont {Thompson}},\ and\ \bibinfo {author}
  {\bibfnamefont {J.~L.}\ \bibnamefont {Sarrao}},\ }\bibfield  {title}
  {\bibinfo {title} {Superconductivity in \textrm{CeCoIn$_{5-x}$Sn$_{x}$}: Veil
  over an ordered state or novel quantum critical point?},\ }\href@noop {}
  {\bibfield  {journal} {\bibinfo  {journal} {Phys. Rev. Lett.}\ }\textbf
  {\bibinfo {volume} {94}},\ \bibinfo {pages} {047001} (\bibinfo {year}
  {2005})}\BibitemShut {NoStop}%
\bibitem [{\citenamefont {Jang}\ \emph {et~al.}(2017)\citenamefont {Jang},
  \citenamefont {Portnichenko}, \citenamefont {Cameron}, \citenamefont
  {Friemel}, \citenamefont {Dukhnenko}, \citenamefont {Shitsevalova},
  \citenamefont {Filipov}, \citenamefont {Schneidewind}, \citenamefont
  {Ivanov}, \citenamefont {Inosov} \emph
  {et~al.}}]{Jang17-Ce1-xLaxB6-Correlations-Effective-mass-Fluctuation}%
  \BibitemOpen
  \bibfield  {author} {\bibinfo {author} {\bibfnamefont {D.}~\bibnamefont
  {Jang}}, \bibinfo {author} {\bibfnamefont {P.~Y.}\ \bibnamefont
  {Portnichenko}}, \bibinfo {author} {\bibfnamefont {A.~S.}\ \bibnamefont
  {Cameron}}, \bibinfo {author} {\bibfnamefont {G.}~\bibnamefont {Friemel}},
  \bibinfo {author} {\bibfnamefont {A.~V.}\ \bibnamefont {Dukhnenko}}, \bibinfo
  {author} {\bibfnamefont {N.~Y.}\ \bibnamefont {Shitsevalova}}, \bibinfo
  {author} {\bibfnamefont {V.~B.}\ \bibnamefont {Filipov}}, \bibinfo {author}
  {\bibfnamefont {A.}~\bibnamefont {Schneidewind}}, \bibinfo {author}
  {\bibfnamefont {A.}~\bibnamefont {Ivanov}}, \bibinfo {author} {\bibfnamefont
  {D.~S.}\ \bibnamefont {Inosov}}, \emph {et~al.},\ }\bibfield  {title}
  {\bibinfo {title} {Large positive correlation between the effective electron
  mass and the multipolar fluctuation in the heavy-fermion metal ce1- x la x
  b6},\ }\href@noop {} {\bibfield  {journal} {\bibinfo  {journal} {npj quantum
  materials}\ }\textbf {\bibinfo {volume} {2}},\ \bibinfo {pages} {62}
  (\bibinfo {year} {2017})}\BibitemShut {NoStop}%
\bibitem [{\citenamefont {Nakamura}\ \emph {et~al.}(2006)\citenamefont
  {Nakamura}, \citenamefont {Endo}, \citenamefont {Yamamoto}, \citenamefont
  {Isshiki}, \citenamefont {Kimura}, \citenamefont {Aoki}, \citenamefont
  {Nojima}, \citenamefont {Otani},\ and\ \citenamefont
  {Kunii}}]{Nakamura06-Ce1-x-LaxB6-NFL-to-FL}%
  \BibitemOpen
  \bibfield  {author} {\bibinfo {author} {\bibfnamefont {S.}~\bibnamefont
  {Nakamura}}, \bibinfo {author} {\bibfnamefont {M.}~\bibnamefont {Endo}},
  \bibinfo {author} {\bibfnamefont {H.}~\bibnamefont {Yamamoto}}, \bibinfo
  {author} {\bibfnamefont {T.}~\bibnamefont {Isshiki}}, \bibinfo {author}
  {\bibfnamefont {N.}~\bibnamefont {Kimura}}, \bibinfo {author} {\bibfnamefont
  {H.}~\bibnamefont {Aoki}}, \bibinfo {author} {\bibfnamefont {T.}~\bibnamefont
  {Nojima}}, \bibinfo {author} {\bibfnamefont {S.}~\bibnamefont {Otani}},\ and\
  \bibinfo {author} {\bibfnamefont {S.}~\bibnamefont {Kunii}},\ }\bibfield
  {title} {\bibinfo {title} {Unusual evolution of the conduction-electron state
  in ${\mathrm{ce}}_{x}{\mathrm{la}}_{1\ensuremath{-}x}{\mathrm{b}}_{6}$ from
  non-fermi liquid to fermi liquid},\ }\href@noop {} {\bibfield  {journal}
  {\bibinfo  {journal} {Phys. Rev. Lett.}\ }\textbf {\bibinfo {volume} {97}},\
  \bibinfo {pages} {237204} (\bibinfo {year} {2006})}\BibitemShut {NoStop}%
\bibitem [{\citenamefont {van~der Marel}\ \emph {et~al.}(2011)\citenamefont
  {van~der Marel}, \citenamefont {van Mechelen},\ and\ \citenamefont
  {Mazin}}]{vanderMarel11-SrTiO3-FL-SUC}%
  \BibitemOpen
  \bibfield  {author} {\bibinfo {author} {\bibfnamefont {D.}~\bibnamefont
  {van~der Marel}}, \bibinfo {author} {\bibfnamefont {J.~L.~M.}\ \bibnamefont
  {van Mechelen}},\ and\ \bibinfo {author} {\bibfnamefont {I.~I.}\ \bibnamefont
  {Mazin}},\ }\bibfield  {title} {\bibinfo {title} {Common {F}ermi-liquid
  origin of ${T}^{2}$ resistivity and superconductivity in $n$-type
  \textrm{SrTiO}$_{3-\delta}$},\ }\href@noop {} {\bibfield  {journal} {\bibinfo
   {journal} {Phys. Rev. B}\ }\textbf {\bibinfo {volume} {84}},\ \bibinfo
  {pages} {205111} (\bibinfo {year} {2011})}\BibitemShut {NoStop}%
\bibitem [{\citenamefont {Jacko}\ \emph {et~al.}(2009)\citenamefont {Jacko},
  \citenamefont {Fjærestad},\ and\ \citenamefont
  {Powell}}]{Jacko09-Kadowaki-Woods-Strongly-correlated-Metals}%
  \BibitemOpen
  \bibfield  {author} {\bibinfo {author} {\bibfnamefont {A.~C.}\ \bibnamefont
  {Jacko}}, \bibinfo {author} {\bibfnamefont {J.~O.}\ \bibnamefont
  {Fjærestad}},\ and\ \bibinfo {author} {\bibfnamefont {B.~J.}\ \bibnamefont
  {Powell}},\ }\bibfield  {title} {\bibinfo {title} {A unified explanation of
  the kadowaki-woods ratio in strongly correlated metals},\ }\href@noop {}
  {\bibfield  {journal} {\bibinfo  {journal} {Nat Phys}\ }\textbf {\bibinfo
  {volume} {5}},\ \bibinfo {pages} {422} (\bibinfo {year} {2009})}\BibitemShut
  {NoStop}%
\end{thebibliography}

\appendix
\renewcommand{\thefigure}{A\arabic{figure}}
\setcounter{figure}{0}

%\newpage
%\beginsupplement
\setcounter{figure}{0} 
\setcounter{table}{0} 
\setcounter{section}{0} 
\setcounter{equation}{0} 
\renewcommand{\thesection}{\Roman{section}} 
\renewcommand{\thesection}{A\arabic{section}}   
\renewcommand{\theequation}{A\arabic{equation}}
\renewcommand{\thetable}{A\arabic{table}}   
\renewcommand{\thefigure}{A\arabic{figure}}
\renewcommand{\thesection}{A\arabic{section}}
\newpage
\section{Distinct contrast between the properties of the FL state of a QCHF superconductor and that of conventional FL superconductor \label{Sec.A-Comparison-FLs}}
Based on the analysis in §\ref{Sec.Emprical-Derivation}, we highlight a striking contrast between the FL state of QCHF superconductors and that of conventional FL superconductors. First, the $A$ coefficient in conventional FL systems is much smaller than the strongly enhanced $A$ observed in QCHF superconductors. More importantly, in weakly coupled conventional FL superconductors, $\rho^{\mbox{\tiny{FL}}}_{\mbox{\tiny{0}}}$ shows no correlation with either $T_c$ or $A$. This is expected, as $\rho^{\mbox{\tiny{FL}}}_{\mbox{\tiny{0}}}$ arises solely from electron-impurity scattering, $\rho^{\mbox{\tiny{FL}}}_{\mbox{\tiny{0}}} \sim |V_{\text{imp}}|^2$, whereas $T_c \sim e^{-1/\lambda}$ depends on the electron-phonon interaction, $\lambda \sim |V_{\text{ep}}|^2$, and $A$ reflects electron-electron interactions, $A \sim |V^{sf}_{ee}|^2$.
Nonetheless, there are two notable exceptions where a correlation between $T_c$ and $A$ has been reported. First, when both quantities can be expressed in terms of Landau parameters that satisfy specific conditions (see Ref.\,\Onlinecite{vanderMarel11-SrTiO3-FL-SUC}). Second, in weakly coupled FL BCS-like superconductors described by $T_c = \theta \exp(-1/VN_{E_F})$, where the Kadowaki–Woods relation $A \propto \gamma^2 \propto [N_{E_F}]^2$ holds, leading to $\ln(T_c/\theta) \propto A^{-1/2}$.

It is important to reemphasize that, while both exceptions yield the same BCS-like correlation between $T_c$ and $A$ (see §\ref{Sec.Emprical-Derivation}), neither case establishes a correlation between $A$ and $\rho^{\mbox{\tiny{FL}}}_{\mbox{\tiny{0}}}$ or between $T_c$ and $\rho^{\mbox{\tiny{FL}}}_{\mbox{\tiny{0}}}$. In fact, the Anderson theorem explicitly rules out the latter.
This makes the empirical findings in Figs.\ref{Fig2-CeCu2Ge2}–\ref{Fig4-CeCoIn5} all the more remarkable. 
%These figures reveal a distinct FL phase within the phase diagrams of QCHF superconductors where superconductivity emerges alongside pressure-dependent correlations among $\rho^{\mbox{\tiny{sf}}}_{\mbox{\tiny{0}}}$, $A$, and $T_c$.
As discussed in the main text, this stark contrast with conventional FL behavior compels us to conclude that a single underlying quantum fluctuation must be responsible for the observed correlations among $T_c$ (a hallmark of superconductivity), $A$ (characteristic of the FL ground state), and $\rho^{\mbox{\tiny{sf}}}_{\mbox{\tiny{0}}}$ (reflecting residual scattering, far greater than in conventional FLs). 
%Such correlated behavior is reminiscent of defect-induced, boson-mediated electron-electron scattering in systems with lattice disorder (see Ref.\,\Onlinecite{21-FL-SC-Defectal-Reconciled}).

%
\section{Derivation of Kadowaki-Woods ratio and gap-to-\protect{$T_c$} ratio of a general fluctuation-bearing superconductor \label{Sec. Kadowaki-Woods-ratio}}
We show below that our theoretical approach is also suitable for deriving further analytic expressions: (i) the Kadowaki-Woods ratio and (ii) the gap-to-$T_c$ ratio, both for a general fluctuation-bearing superconductor.

The Kadowaki-Woods ratio is usually defined as $A/\gamma^2$ which is expected to be a universal 
constant in Fermi liquids since $A\propto {m^{\star}}^2$ and $\gamma\propto m^{\star}$. However, it was observed that ratios of the heavy-Fermions are widely different from those of the transition metals compounds; in fact, ratios within the very same class may differ by order of magnitude.
Jacko \textit{et al.}\cite{Jacko09-Kadowaki-Woods-Strongly-correlated-Metals} accounted for such a difference, among a variety of strongly-coupled systems, by demonstrating that
\begin{equation}
\frac{A}{\gamma^2}= \Big( \frac{81}{4\pi\hbar k_B^2 e^2}\Big) \Big(\frac{1}{d^2 n N^2(\epsilon_F)\langle  v_{0x}^2\rangle}\Big)= \Big(f_{con}\Big)\Big( \frac{1}{f_{mat}}\Big),
\label{Eq.A-KDW-ratio}
\end{equation}
where $\langle v_{0x}^2\rangle$ is a Fermi surface average of the carrier velocity squared
that accounts for anisotropies, $e$ is the electric charge of the direct, Coulomb, electric-electric
interaction, $n$ is the carrier density, $N(\epsilon_F)$ is the density of states at the Fermi level, 
and $d \sim 1$ is a dimensionless number. The $f_{con}$ factor contains fundamental constants while $f_{mat}$ contains material-dependent parameters. Apparently, in contrast to the universal $\frac{A\, f_{mat}}{\gamma^2}$ ratio, $\frac{A}{\gamma^2}$ of Eq.\ref{Eq.A-KDW-ratio} is material-dependent: depriving the Kadowaki-Woods ratio from its universal character.
This shortcoming becomes most evident within the FL regime of the QCHF superconductors (see Figs.\ref{Fig2-CeCu2Ge2}-\ref{Fig4-CeCoIn5}); in this regime, our evaluation of the Kadowaki-Woods ratio (Eq.\ref{Eq.A-KDW-ratio}) gives  
\begin{equation}
\frac{A}{\gamma^2}=\Big(\frac{81}{4\pi\hbar k_B^2 e^2}\Big)\Big(
\frac{F_{\ell}}{d^2 n N^2(\epsilon_F)\langle v_{0x}^2\rangle}\Big)=\Big(f_{con}\Big)\Big( \frac{F_{\ell}}{f_{mat}}\Big) ,
\label{Eq.A-KDW-ratio-Defectal}
\end{equation}
%frac{A(\ell)}{\gamma^2}=\frac{81}{4\pi\hbar k_B^2 e^2}\left(\frac{F_\ell}{F_\infty}\right)
wherein the material-dependent $F_{\ell}$ factor is as described in Eq.\ref{Eq.A-vs-Lambda}. It is not close to 1 as assumed during the derivation of Eq.\ref{Eq.A-KDW-ratio}.\cite{Jacko09-Kadowaki-Woods-Strongly-correlated-Metals} Rather $F_{\ell} > 1$: a larger \textit{apparent} ratio because of the easing of the kinematic constraints in these spin-fluctuation-bearing systems. Then, $\frac{A}{\gamma^2}\Big( \frac{f_{mat}}{F_{\ell}}\Big)=\frac{A\,f_{mat}^{'}}{\gamma^2}=f_{con}$ is universal, as in Ref.\,\Onlinecite{Jacko09-Kadowaki-Woods-Strongly-correlated-Metals}. 

We also calculated the gap-to-$T_c$ ratio  for such QCHF superconductors, beyond the $\theta/T_c\rightarrow\infty$ approximations:
\begin{equation}
\frac{2\Delta(\ell)}{k_B T_c(\ell)}=3.53\left\{1+12.5 \left[\frac{T_c(\ell)}{\theta}\right]\ln{\left[\frac{\theta}{2T_c(\ell)}\right]}\right\}.
\label{Eq.A-gap-to-Tc-ratio}
\end{equation}
Thus, the gap-to-$T_c$ ratio can be fine-tuned (from $2\Delta(\infty)/k_B T_c(\infty)=3.53$, the universal BCS value, up to a higher nonuniversal value) by varying the pressure or any control-parameter that modifies $\ell$. 

\end{document}